\documentclass[fleqn,usenatbib,usedcolumn]{mnras}
\usepackage[british]{babel}             % British English hyphenation
\usepackage{txfonts}                  % Good fonts
\let\la=\lesssim % for less than similar from newtxmath, not \la from mnras.cls
\usepackage[T1]{fontenc}                % Good font encoding
\usepackage{graphicx}                   % Including figures
\usepackage{rotating}
\usepackage{dcolumn}
\newcolumntype{d}{D{.}{.}{-1}}
\def\dhead#1{\multicolumn{1}{c}{#1}}
\def\twolines#1#2{$\kern-6pt\Big\{ {\textrm{#1\hfill}\atop\textrm{#2\hfill}}$}
\def\vpad{{\Large$\mathstrut$}}
\usepackage{multirow}
\usepackage{hyperref}
\graphicspath{{.}{./FIGS/}}

\hypersetup{pdfauthor={I. H. Whittam, J. M. Riley, D. A. Green and M. J. Jarvis},
               pdftitle={The faint radio source population at 15.7~GHz -- III. A high-frequency study of HERGs and LERGs},
               pdfkeywords={galaxies: active, radio continuum: galaxies, catalogues, surveys},
               bookmarksnumbered=true}
\hypersetup{colorlinks=true,
            linkcolor=blue,
            citecolor=blue,
            filecolor=blue,
            urlcolor=blue}
\volume{{\rm in prep.}}
\title[The faint source population at 15.7~GHz -- III. A high-frequency study of HERGs and LERGs]{The faint source population at 15.7~GHz -- III. A high-frequency study of HERGs and LERGs}

\author[I.~H.~Whittam et al.]{I.~H.~Whittam$^{1}$\thanks{email:
\texttt{iwhittam@uwc.ac.za}}, J.~M.~Riley$^2$, D.~A.~Green$^2$ and M. J. Jarvis$^{1,3}$\\
   $^{1}$Physics and Astronomy Department, University of the Western Cape, Bellville 7535, South Africa\\
   $^{2}$Astrophysics Group, Cavendish Laboratory, 19 J.~J.~Thomson Avenue, Cambridge CB3 0HE\\
   $^{3}$Astrophysics, University of Oxford, Denys Wilkinson Building, Keble Road, Oxford, OX1 3RH}

\date{Accepted ---; received ---; in original form ---}

\pagerange{\pageref{firstpage}--\pageref{lastpage}}

\pubyear{2016}
\begin{document}

\label{firstpage}

\maketitle

\begin{abstract}
A complete sample of 96 faint ($S > 0.5$~mJy) radio galaxies is selected from the Tenth Cambridge (10C) survey at 15.7~GHz. Optical spectra are used to classify 17 of the sources as high-excitation or low-excitation radio galaxies (HERGs and LERGs respectively), for the remaining sources three other methods are used; these are optical compactness, X-ray observations and mid-infrared colour--colour diagrams. 32 sources are HERGs and 35 are LERGs while the remaining 29 sources could not be classified. We find that the 10C HERGs tend to have higher 15.7-GHz flux densities, flatter spectra, smaller linear sizes and be found at higher redshifts than the LERGs. This suggests that the 10C HERGs are more core dominated than the LERGs.

Lower-frequency radio images, linear sizes and spectral indices are used to classify the sources according to their radio morphology; 18 are Fanaroff and Riley type I or II sources, a further 13 show some extended emission, and the remaining 65 sources are compact and are referred to as FR0 sources. The FR0 sources are sub-divided into compact, steep-spectrum (CSS) sources (13 sources) or GHz-peaked spectrum (GPS) sources (10 sources) with the remaining 42 in an unclassified class. FR0 sources are more dominant in the subset of sources with 15.7-GHz flux densities $<$1~mJy, consistent with the previous result that the fainter 10C sources have flatter radio spectra. 

The properties of the 10C sources are compared to the higher-flux density Australia Telescope 20~GHz (AT20G) survey. The 10C sources are found at similar redshifts to the AT20G sources but have lower luminosities. The nature of the high-frequency selected objects change as flux density decreases; at high flux densities the objects are primarily quasars, while at low flux densities radio galaxies dominate.
\end{abstract}


%------------------------------------------------------------------------------%

%@arxiver{{HERG_z-crop.pdf,{HERG_alpha-crop.pdf,Mahony-eps-converted-to.pdf}

\section{Introduction}\label{section:intro}

In the first two papers in this series (\citealt{2013MNRAS.429.2080W,Paper II}, referred to as Papers I and II respectively) we studied the properties of a sample of sources selected from the Tenth Cambridge (10C; \citealt{2011MNRAS.415.2708D,2011MNRAS.415.2699F}) survey at 15.7 GHz in the Lockman Hole. The 10C survey is complete to 0.5~mJy, making it the deepest high-frequency radio survey to date. In Paper II we found that the vast majority ($\geqslant 94$ per cent) of the sources in the 10C sample are radio galaxies; the 10C sample is therefore the ideal starting point for a study of the properties of faint radio galaxies.

It is well known that powerful extended radio galaxies can be split into two classes according to their morphology following the Fanaroff--Riley (FR) scheme; FRI sources have the highest surface brightness near the core, while FRII sources are brightest near the lobe edges and have more collimated jets \citep{1974MNRAS.167P..31F}. There is also a divide in luminosity between the two classes, with the more powerful FRII sources having $L_{1.4~\rm GHz} > 10^{24.5}~\rm W \, Hz^{-1}$, while the FRI sources predominantly have luminosities below this value. Some recent papers (e.g.\ \citealt{2014MNRAS.438..796S}, \citealt{2015A&A...576A..38B}) have introduced the `FR0' classification to describe radio galaxies which lack the extended emission typical of FRI and FRII sources. There are two main types of compact radio galaxies; compact-steep spectrum sources (CSS), which have linear sizes between 1 and 15~kpc and convex spectra which peak below $\sim 500$~MHz, and gigahertz-peaked spectrum sources (GPS), which are smaller with linear sizes $\lesssim 1$~kpc and have spectra which peak between 500~MHz and 10~GHz (see \citealt{1998PASP..110..493O} for a review). Both classes of compact sources are powerful, with typical luminosities $L_{1.4~\rm GHz} \gtrsim 10^{25}~\rm W \, Hz^{-1}$. There is evidence that both GPS and CSS sources are young radio galaxies (\citealt{2008arXiv0802.1976S} and references therein). 

It has been known for some time \citep{1979MNRAS.188..111H} that the properties of radio galaxies are not fully explained by the conventional model of an AGN, consisting of an accretion disk surrounded by a dusty torus \citep{1993ARA&A..31..473A}. Based on this model, we would expect radio-loud objects viewed close to the jet axis to show both broad and narrow optical emission lines, while radio-loud objects viewed at larger angles to the jet would only show narrow lines and would have a clear mid-infrared signature of the dusty torus. However, many radio-loud AGN lack the expected narrow-line optical emission and do not display evidence of an obscuring torus (e.g.\ \citealt{2004ApJ...602..116W}).

Subsequent studies have suggested that there are two fundamentally distinct accretion modes, known as `cold mode' and `hot mode' (see \citealt{2005MNRAS.362...25B,2007MNRAS.376.1849H}) which could be responsible for these differences (these modes are sometimes referred to as `quasar' and `radio' modes respectively). Cold-mode accretion occurs when cold gas is accreted onto the central black hole through a radiatively efficient, geometrically thin, optically thick accretion disk (e.g.\ \citealt{1973A&A....24..337S}) and gives rise to the traditional model of an AGN \citep{1993ARA&A..31..473A}. These objects therefore show high-excitation lines in their optical spectra, so are often referred to as high-excitation radio galaxies (HERGs). `Hot mode' sources, however, are thought to be fuelled by the accretion of warm gas through advection-dominated accretion flows (e.g.\ \citealt{1995ApJ...452..710N}) and lack many of the typical signatures of AGN, such as strong optical emission lines. These objects are often referred to as low-excitation radio galaxies (LERGs). LERGs typically show no evidence for a dusty torus \citep{2006ApJ...647..161O} or for accretion-related X-ray emission \citep{2006MNRAS.370.1893H}. There are also differences in the host galaxies of the two populations, with HERGs found in less massive and bluer galaxies than LERGs \citep{2005MNRAS.362....9B,2008A&A...490..893T,2010MNRAS.406.1841H,2011MNRAS.410.1360H,2015MNRAS.450.1538W}. Although both HERGs and LERGs are found across the full range in radio luminosities, LERGs seem to dominate at lower luminosities and HERGs at higher luminosities \citep{2012MNRAS.421.1569B}. There is a substantial overlap between the FRI/II classification and the HERG and LERG classes, with most FRI sources being LERGs and most FRIIs being HERGs \citep{2012MNRAS.421.1569B}. There are, however, differences in the two classifications, in particular a significant population of FRII LERGs has been found \citep{1994ASPC...54..201L}.

\citet{2012MNRAS.421.1569B} showed that low-redshift HERGs and LERGs have distinct accretion rates; HERGs typically accrete at between 1 and 10 per cent of their Eddington rate, while LERGs generally have accretion rates much less than 1 per cent of their Eddington rate. \citet{2015MNRAS.447.1184F} also find evidence for this at $z \sim 1$ using mid-infrared data. Therefore a picture is emerging where HERGs accrete cold gas at a relatively high rate \citep{2012MNRAS.421.1569B} and radiate efficiently across the whole electromagnetic spectrum (e.g.\ \citealt{1994ApJS...95....1E}). This causes them to produce a stable accretion disc and therefore display the typical properties of an AGN. The cold gas leads to star formation \citep{2013MNRAS.432..609V,2013MNRAS.429.2407H}, causing the host galaxies of HERGs to be relatively blue (e.g.\ \citealt{2003MNRAS.346.1055K}). In addition, HERGs are more prevalent at earlier cosmic epochs, where higher rates of mergers and interactions provided a steady supply of cold gas. 

LERGs, however, slowly accrete warm gas from the X-ray emitting halo of the galaxy or cluster. They radiate inefficiently, emitting the bulk of their energy in kinetic form as powerful jets (e.g.\ \citealt{2007MNRAS.381..589M}). They therefore tend to be hosted by massive galaxies, often at the centres of groups or clusters \citep{1990AJ.....99...14B,2007MNRAS.379..894B}. These galaxies have an old, passive stellar population and show little cosmic evolution out to $z \sim 1$ \citep{2004MNRAS.352..909C,2007MNRAS.381..211S,2014MNRAS.445..955B}.

The fundamentally different accretion modes of HERGs and LERGs cause them to have different radio properties. \citet{2011MNRAS.417.2651M} studied the properties of high flux density ($S_{20~\rm GHz} > 40$~mJy) HERGs and LERGs in the Australia Telescope 20~GHz (AT20G; \citealt{2010MNRAS.402.2403M}) sample and found that while both accretion modes display a range of radio properties, a higher fraction of HERGs are extended and have steep spectra. They suggest that this is because HERGs are accreting more efficiently than LERGs, and therefore have a greater chance of producing more luminous jets and lobes. They also find that HERGs display different properties depending on their orientation, with objects displaying broad emission lines tending to be flat spectrum, and objects with narrow lines tending to be steep spectrum, as predicted by orientation models (\citealt{1993ARA&A..31..473A,1995PASP..107..803U}). LERGs, however, display no orientation effects. These results are consistent with the scenario in which HERGs have a typical AGN accretion disk and torus while LERGs do not. 

Most studies of HERGs and LERGs have focussed on high flux density (e.g.\ \citealt{2011MNRAS.417.2651M}) or low redshift (e.g.\ \citealt{2012MNRAS.421.1569B}) sources, with the exception of \citet{2015MNRAS.447.1184F}, who studied a small sample at $z \sim 1$. In this work, we focus on a sample of high-frequency selected mJy and sub-mJy sources. In Section~\ref{section:radio-data} we outline the data used in this study. In Section~\ref{section:classifications} we describe how this data is used to separate the radio galaxies into HERGs and LERGs and the properties of these HERGs and LERGs are explored in Section~\ref{section:object_properties}. In Section~\ref{section:FR-sources} the sample is split into different radio morphological classes and the properties of these classes are discussed and compared to the HERG and LERGs classifications. The sample is compared to the higher-flux density AT20G sample in Section~\ref{section:other-work} and the conclusions are presented in Section~\ref{section:chap5-conclusions}.

\section{Sample definition and data used}\label{section:radio-data}

The 10C survey was observed with the Arcminute Microkelvin Imager (AMI; \citealt{2008MNRAS.391.1545Z}) and covered a total of 27~deg$^2$ complete to 1~mJy across ten different fields. A further 12~deg$^2$, contained within these fields, is complete to 0.5~mJy. In Paper I we selected a sample of 296 sources from two fields of the 10C survey in the Lockman Hole. This sample was matched to a number of lower-frequency (and higher resolution) catalogues available in the field; a deep 610~MHz Giant Meterwave Radio Telescope (GMRT) image \citep{2008MNRAS.387.1037G,2010BASI...38..103G}, a 1.4~GHz Westerbork Synthesis Radio Telescope (WSRT) image \citep{2012rsri.confE..22G}, two deep Very Large Array (VLA) images at 1.4~GHz which only cover part of the field \citep{2006MNRAS.371..963B,2008AJ....136.1889O}, the National Radio Astronomy Observatory (NRAO) VLA Sky Survey (NVSS; \citealt{1998AJ....115.1693C}) and the Faint Images of the Radio Sky at Twenty centimetres (FIRST; \citealt{1997ApJ...475..479W}). For full details of the catalogues used and the matching procedure see Paper I. These data enabled spectral indices and angular sizes to be found for the 10C sources.

In Paper II we selected a complete sub-sample of 96 sources from the full 10C sample used in Paper I. This sub-sample is the subject of the remainder of this paper; it is complete to 0.5~mJy at 15.7~GHz and is in a region of the field with particularly good low-frequency (610~MHz and 1.4~GHz) coverage. All but one of the sources have counterparts in at least one of the lower-frequency catalogues so spectral indices and accurate positions are available. This sample was matched to the `Data Fusion' catalogue, which was compiled by \citet{2015arXiv150806444V}\footnote{\texttt{http://www.mattiavaccari.net/df/}} and contains most of the multi-wavelength data available in the field. The catalogues included are: the \emph{Spitzer} Wide-Area Infrared Extragalactic survey (SWIRE; see \citealt{2003PASP..115..897L}), the \emph{Spitzer} Extragalactic Representative Volume Survey (SERVS; see \citealt{2012PASP..124..714M}) and the United Kingdom Infrared Telescope (UKIRT) Infrared Deep Sky Survey (UKIDSS; see \citealt{2007MNRAS.379.1599L}) with deep optical photometric data taken by \citet{2011MNRAS.416..927G} (GS11) at the Isaac Newton Telescope (INT) and the Kitt Peak National Observatory (KPNO). In total 80 out of the 96 sources have at least one multi-wavelength counterpart. Spectroscopic redshifts are available for 24 sources and photometric redshifts were estimated for the remaining sources where possible, giving redshift values or estimates for 77 out of the 96 sources. The median redshift of the sample is 0.91. In Paper II we used these multi-wavelength data to show that $\geqslant 94$ per cent of sources in this sample are radio galaxies, with a very small number of low redshift starforming galaxies. For further details of the multi-wavelength data, the methods used to match the catalogues and compute photometric redshifts see Paper II. 

\begin{figure}
\centerline{\includegraphics[width=10cm,angle=270]{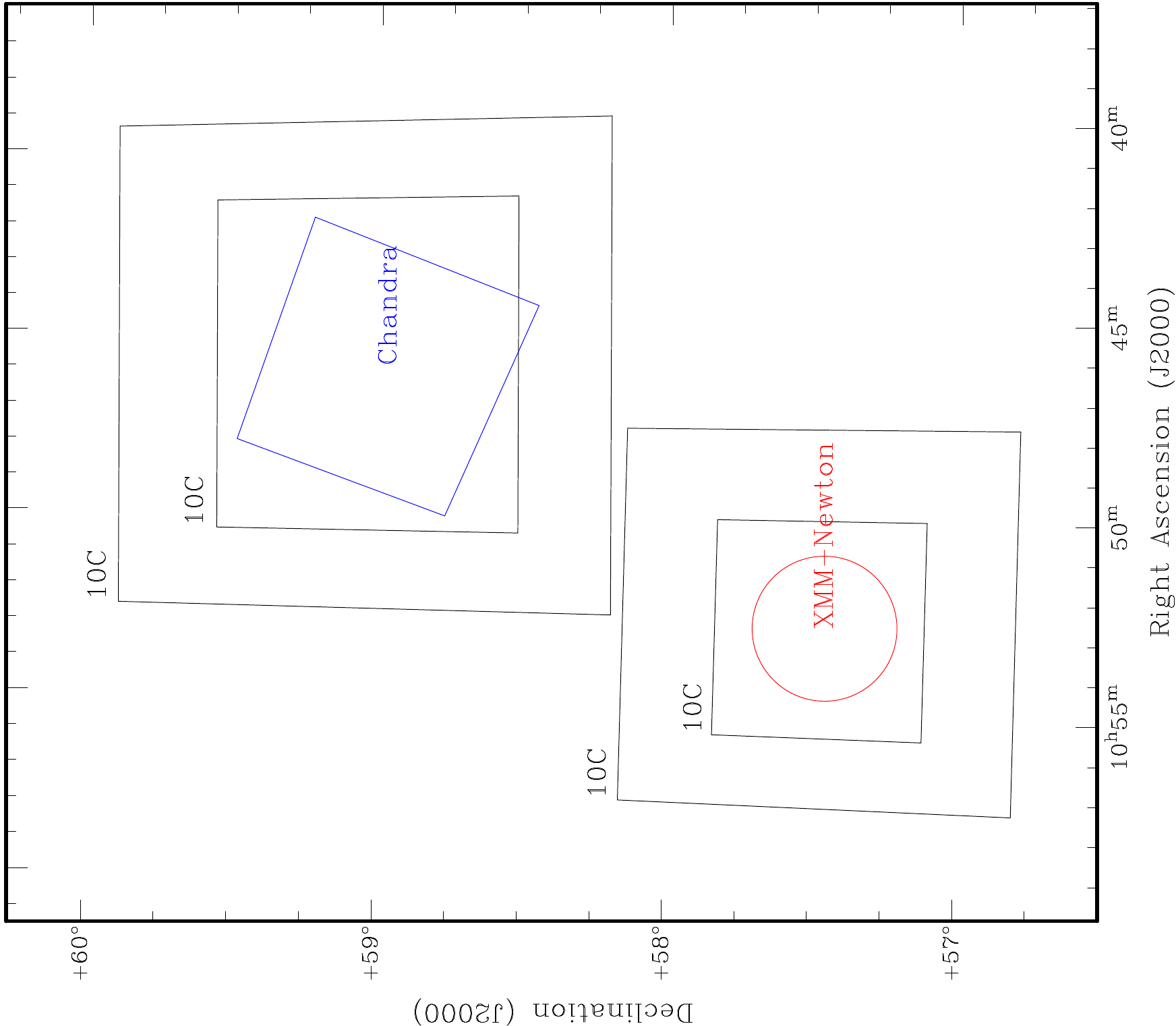}}
\caption{Positions of the two X-ray surveys in the Lockman Hole.  The \citet{2009ApJS..185..433W} \emph{Chandra} survey is shown in blue and the \citet{2008A&A...479..283B} \emph{XMM-Newton} survey is shown in red. Positions of the 10C radio survey fields are shown in black for reference (the large rectangles indicate the regions complete to 1~mJy and small rectangles contained within show the regions complete to 0.5~mJy). }\label{fig:xray}
\end{figure}

In addition, to help elucidate the nature of these radio galaxies, X-ray data are used in this paper. There are two separate X-ray survey fields in the Lockman Hole 10C survey area: one field is observed by \emph{Chandra} \citep{2009ApJS..185..433W} and one by \emph{XMM-Newton} \citep{2008A&A...479..283B}. Fig.~\ref{fig:xray} illustrates the positions of these two surveys. The \emph{Chandra} Lockman Hole field covers 0.7~deg$^2$ and detected 775 X-ray sources to a limiting broadband (0.3 to 8~keV) flux $\sim4 \times 10^{-16}~\rm{erg ~cm^{-2} ~s^{-1}}$. The \emph{XMM-Newton} field covers $\sim$0.2~deg$^2$ and detects 409 sources with a sensitivity limit of 1.9, 9 and 180 $\times 10^{-16}~\rm{erg ~cm^{-2} ~s^{-1}}$ in the 0.5 to 2.0, 2.0 to 10.0 and 5.0 to 10.0~keV bands respectively. 

The two X-ray catalogues were matched to the 10C sources using a 5 arcsec match radius (a procedure similar to that described in Section 3.2 in Paper II was carried out to determine an appropriate match radius). For those 10C sources with a match to the Data Fusion catalogue, the SERVS position from the Data Fusion catalogue was used. For the remaining sources the radio position was used, with the radio position chosen in the following order of preference: FIRST, GMRT, BI2006/OM2008, WSRT, 10C.
The results of this matching are summarised in Table \ref{tab:xray-matches}. In total, out of 32 10C sources in the two X-ray survey areas, 15 (47 per cent) have an X-ray counterpart.

\begin{table}
\caption{A summary of the X-ray counterparts found for 10C sources.}\label{tab:xray-matches}
\medskip
\centering
\begin{tabular}{llll}\hline
 Field & No. of 10C     & No. with an  & Percentage with\\
       & sources in field  & X-ray match     & an X-ray match\\\hline
Chandra    & 19 & 7 & 37\\
XMM-Newton & 13 & 8 & 62\\\hline
\end{tabular} 
\end{table}

%----------------------------------------------------------------------------------------------------%
\section{Classifying HERGs and LERGs}\label{section:classifications}

Three different methods are used to classify the radio sources as either HERGs or LERGs in this paper. These are described in this section, then an overall classification scheme is presented.

\subsection{Optical compactness}\label{section:opt-compact}

The compactness of an object in the optical image can help to indicate whether it is a HERG or a LERG. For example, \citet{2011MNRAS.417.2651M} compared optical compactness classifications from the superCOSMOS survey \citep{2001MNRAS.326.1279H} to spectral classifications and found that most of the objects which displayed broad emission lines, and are therefore HERGs, were point-like in the optical. However, they found that 27 per cent of the objects which were extended in the optical also had broad emission lines. This suggests that if an object is point-like it is a HERG in the form of a quasar, however, an object which is extended in the optical is not necessarily a LERG, in line with orientation effects.

Optical compactness information is contained in the full optical catalogue produced by \citet{2011MNRAS.416..927G} (details of which are given in Section 2.2 in Paper II). This catalogue contains optical compactness classification flags in each band ($r,~g,~i ~{\rm and} ~z$), which are then combined into a merged probability of each object being point-like or extended (`pstar' and `pgalaxy' respectively). Any source with pgalaxy $> 90$~per cent was considered extended, and any source with pstar $> 90$~per cent was considered point-like. 

Optical compactness information is available for 59 of the sources and all 59 fell into one or other of the two categories: nine are classified as point-like, and are therefore probably HERGs, while the remaining 50 are classified as extended, and therefore could be either HERGs or LERGs, so remain unclassified. The classifications are summarised in Table~\ref{tab:obj_properties}.

\subsection{Mid-infrared colours}\label{section:colour-colour}

Mid-infrared colour--colour diagrams are an effective way of separating HERGs and LERGs. The characteristic signature of a dusty AGN torus, typically present in HERGs but missing from LERGs, is power-law continuum emission so these objects fall in a particular region on these mid-infrared colour--colour diagrams \citep{2004ApJ...602..116W,2005Natur.436..666M,2006ApJ...647..161O}. 

\citet{2004ApJS..154..166L} defined a region in the mid-infrared colour--colour diagram using \emph{Spitzer} IRAC filters in which objects with power-law continuum emission (produced by a dusty AGN torus) are expected to lie. HERGs are expected to lie inside this `AGN region', while LERGs are not.

\begin{figure}
\centerline{\includegraphics[width=\columnwidth]{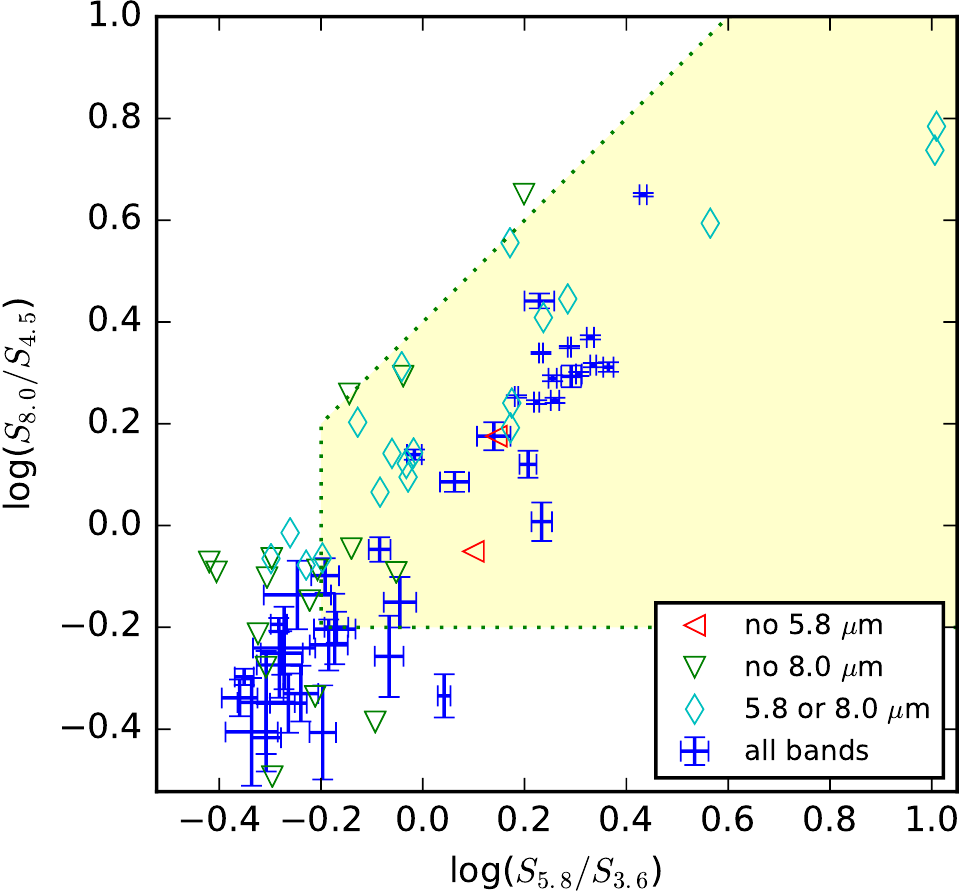}}
\caption{ A mid-infrared colour--colour diagram, including upper limits for sources which are not detected in one or two of the mid-infrared bands (only the 76 sources which are detected in at least two mid-infrared band are shown). Error bars are shown on points detected in all four bands. Sources without a 5.8~$\muup \rm m$ detection could move to the left, sources without an 8.0~$\muup \rm m$ detection could move down, and sources without 5.8 or 8.0~$\muup \rm m$ values could move left and/or down. The region inside the dotted lines (shaded yellow in the online version) is the \citeauthor{2004ApJS..154..166L} AGN area, in which HERGs are expected to lie. }\label{fig:Lacy_errors}
\end{figure}

\citet{2012ApJ...748..142D} further investigated mid-infrared selection of AGN and defined a narrower wedge, which has significantly lower contamination from starforming sources than the \citet{2004ApJS..154..166L} area but consequently misses some AGN. In this paper we use the area defined in \citet{2004ApJS..154..166L,2007AJ....133..186L}, as we know that there are essentially no starforming sources in our sample so this provides the most reliable way of selecting all sources with an AGN torus. This region is defined as follows:
\begin{eqnarray}
{\rm log}_{10}(S_{8.0}/S_{4.5}) &\leq& 0.8 {\rm log}_{10}(S_{5.8}/S_{3.6}) + 0.5,\nonumber\\
{\rm and~~} {\rm log}_{10}(S_{5.8}/S_{3.6}) &>& -0.2,\nonumber\\
{\rm and~~} {\rm log}_{10}(S_{8.0}/S_{4.5}) &>& -0.2,\label{eqn:lacy}
\end{eqnarray}
where $S_x$ is the flux density in the $x~\muup$m \emph{Spitzer} band. We will refer to this as the `\citeauthor{2004ApJS..154..166L} AGN area', and classify radio galaxies with host galaxies lying inside this region as HERGs (Fig.~\ref{fig:Lacy_errors}).

Sources dominated by old, red starlight are expected to lie outside the bottom left corner of the \citeauthor{2004ApJS..154..166L} AGN area \citep{2005ApJ...621..256S}. LERGs are generally hosted by old, red elliptical galaxies and are therefore expected to lie in this region.  \citet{2011ApJ...740...20P} studied the properties of 256 faint ($S_{1.4~\rm GHz} > 43~\muup \rm Jy$) radio sources observed at 1.4 and 5~GHz with the VLA in the Chandra Deep Field South using the wealth of ancillary data available in the field. They plotted the positions of four known FRI sources and found that all four lay in the old starlight dominated region (outside the bottom left corner of the \citeauthor{2004ApJS..154..166L} AGN area), consistent with the fact that FRI sources (which are mostly LERGs) are predominantly hosted by red ellipticals (e.g.\ \citealt{2014MNRAS.438.1149G,2013MNRAS.436.3759B}). 

%Low-redshift starforming galaxies are expected to be found in the vertical strip ${\rm log}(S_{5.8}/S_{3.6}) \sim -0.2$ to $-0.8$ and ${\rm log}(S_{8.0}/S_{4.5}) > 0$ \citep{2005ApJ...621..256S}. We therefore do not expect to find 10C sources in this region.

The 10C sources detected in at least two mid-infrared bands are plotted on a mid-infrared colour--colour diagram in Fig.~\ref{fig:Lacy_errors}, where the \citeauthor{2004ApJS..154..166L} AGN area (described by equation \ref{eqn:lacy}) is shown by the shaded region bounded by dotted lines. There are 39 objects detected in all four bands in this diagram and 22 of these lie inside the AGN area. Error bars are included on this figure but are omitted from all further mid-infrared colour--colour diagrams for clarity. 

Fig.~\ref{fig:Lacy_errors} shows that all the 10C sources lie either in the \citeauthor{2004ApJS..154..166L} AGN area, consistent with having a power-law SED, or in the region where objects dominated by old starlight are expected to lie (outside the bottom left corner of the \citeauthor{2004ApJS..154..166L} AGN area). There are no objects in the vertical strip ${\rm log}(S_{5.8}/S_{3.6}) \sim -0.2$ to $-0.8$ and ${\rm log}(S_{8.0}/S_{4.5}) > 0$ where low-redshift ($z<0.5$) young starforming objects are expected to lie \citep{2005ApJ...621..256S}. This supports the conclusion of Paper II that there are no starforming sources in the 10C sample.

It is possible to include upper limits for those sources which are not detected in one or two of the mid-infrared bands by using the 95 per cent completeness limit in each band as an upper limit on the flux density of any source not detected in that band. This provides information about a further 37 objects, giving 76 objects in total (six additional sources are only detected in one band and therefore cannot be placed on the diagram). Fig.~\ref{fig:Lacy_errors} shows the mid-infrared colour--colour diagram for these 76 objects, including upper limits where appropriate. In addition to the 22 objects definitely inside the AGN area, there are 21 objects apparently located inside this region; however the nature of their limits mean that they could move outside it, so their classification is uncertain. Thirty-one objects definitely lie outside this area, 17 with detections in all four mid-infrared bands and 14 with upper limits such that they could not move into the \citeauthor{2004ApJS..154..166L} AGN area; all 31 are therefore classified as LERGs. There are two objects outside the \citeauthor{2004ApJS..154..166L} AGN area area which could move into it. Thus in total there are 23 objects on the plot which may, or may not, lie inside the \citeauthor{2004ApJS..154..166L} AGN area. The classifications are summarised in Table~\ref{tab:obj_properties}.

\subsection{X-ray emission}\label{section:x-ray}

HERGs typically produce X-ray emission, caused by the accretion of matter onto the central black hole. This accretion-related X-ray emission is generally missing from LERGs \citep{2006ApJ...642...96E,2006MNRAS.370.1893H}. X-ray observations are therefore a useful diagnostic of radio source type.

A total of 32 10C sources in this sample are located in either the \emph{Chandra} or the \emph{XMM-Newton} X-ray survey fields. Out of these 32 sources, 15 have an X-ray counterpart (47 percent), and are therefore probably HERGs. This does not rule out the possibility that the remaining objects host an AGN which does not produce bright enough X-ray emission to be detected in these surveys, so we are unable to classify the other 17 sources at this stage. The classifications are summarised in Table~\ref{tab:obj_properties}.

\subsection{Comparing the different methods}\label{section:types_comparison}

\begin{figure*}
\centerline{\includegraphics[width=\columnwidth]{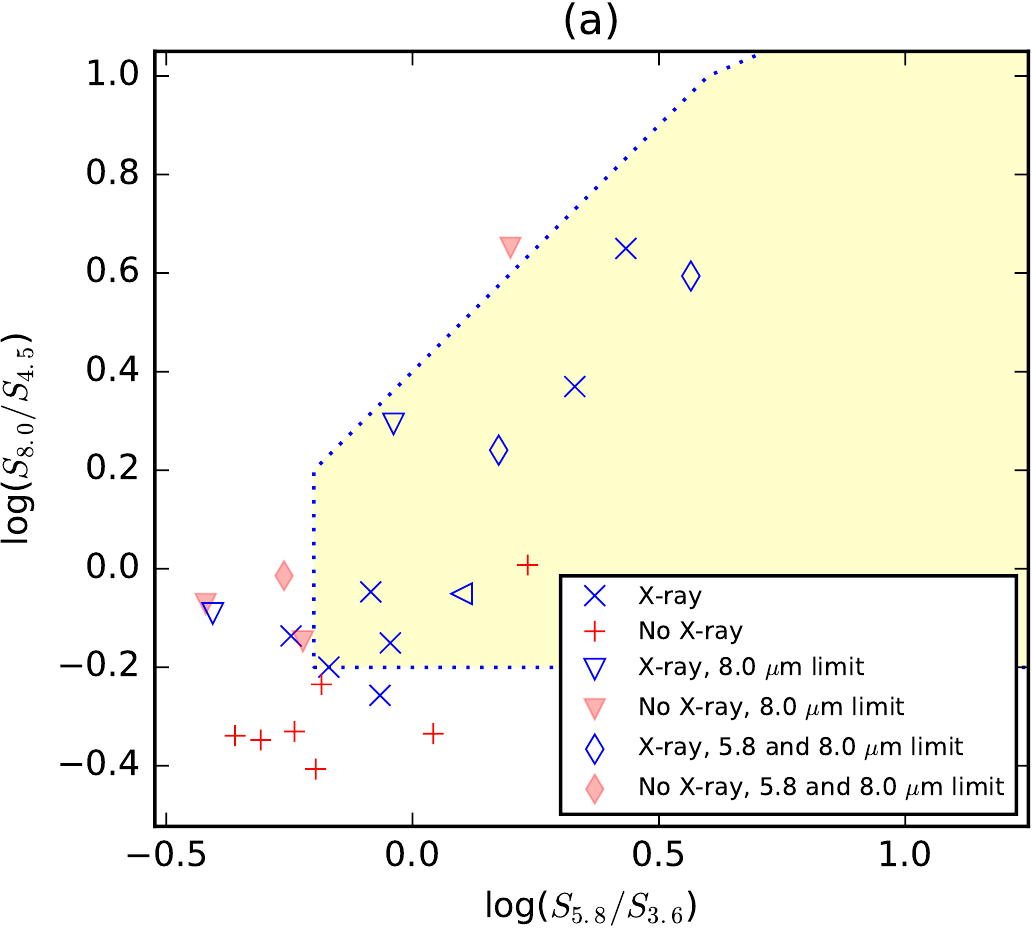}
            \quad
            \includegraphics[width=\columnwidth]{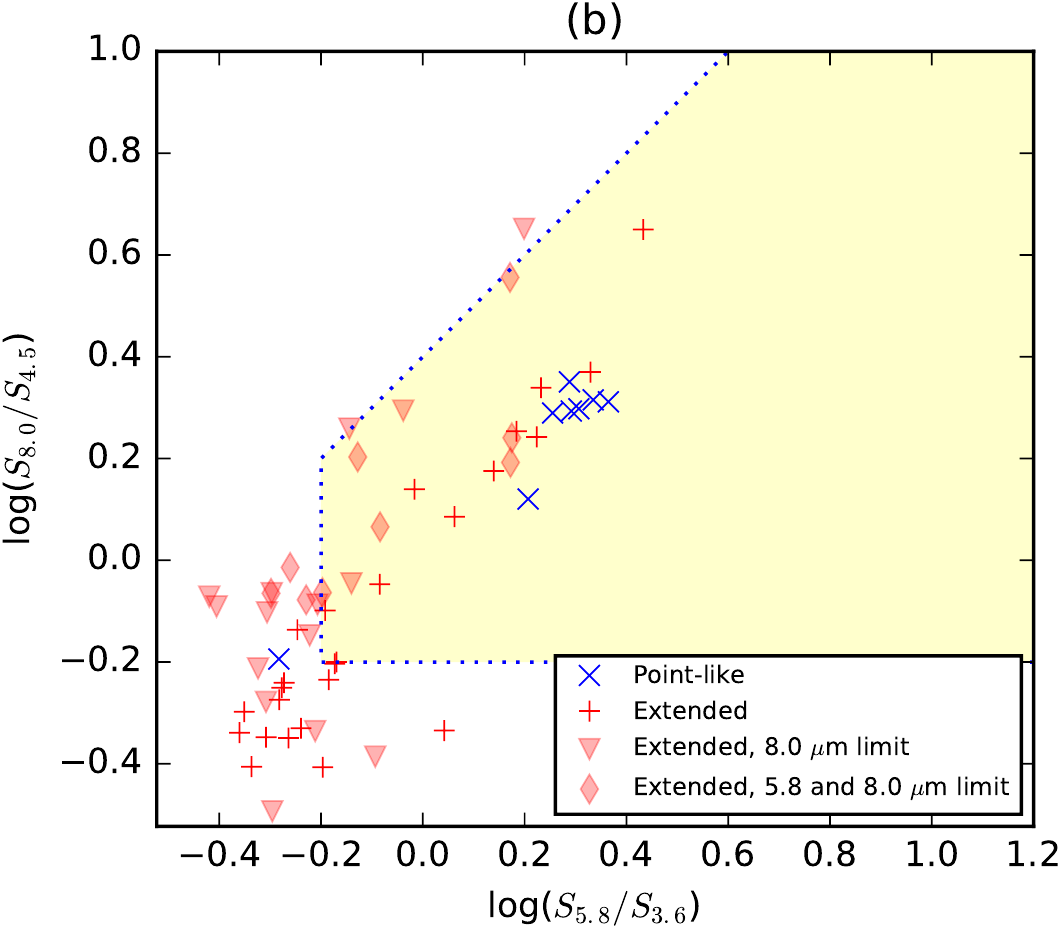}}
\caption{Mid-infrared colour--colour diagrams for sources classified using other parameters. The region inside the dotted lines (shaded yellow in the online version) is the \citeauthor{2004ApJS..154..166L} AGN area, in which HERGs are expected to lie. The left panel shows sources with and without X-ray detections (the 23 sources which are inside the X-ray survey areas and have a detection in at least one mid-infrared band are included); the right panel shows sources split according to their optical compactness (the 55 sources with optical compactness information and a detection in at least two mid-infrared band are included). Error bars are omitted from this plot for clarity, but are included in Fig.~\ref{fig:Lacy_errors}. }\label{fig:Lacy}
\end{figure*}

The mid-infrared separation of HERGs and LERGs is compared to the optical compactness and X-ray emission properties in Fig.~\ref{fig:Lacy}.  Sources which have and have not been detected by X-ray observations are shown in panel (a). All but one of the sources not detected in the X-ray lie outside the \citeauthor{2004ApJS..154..166L} AGN area and five of the seven X-ray-detected sources lie inside the \citeauthor{2004ApJS..154..166L} AGN area with the two remaining sources lying close to the boundary. This shows that these two methods of classifying HERGs and LERGs are generally consistent.

Panel (b) of Fig.~\ref{fig:Lacy} shows sources classified according to their optical compactness on the mid-infrared colour--colour diagram. All but one of the eight sources which are point-like in the optical lie close together in the top right of the \citeauthor{2004ApJS..154..166L} AGN area on this diagram where quasars are expected to lie, showing that there is good agreement between the two methods of classifying radio galaxies. However, 12 sources which are extended in the optical also lie inside the \citeauthor{2004ApJS..154..166L} AGN area on this diagram. This agrees with the findings of \citet{2011MNRAS.417.2651M} that while the majority of point-like sources are HERGs, not all HERGs are point-like.

Twenty sources are in an X-ray survey area and have information about their optical compactness available. Eight out of these 20 sources are detected in the X-ray, suggesting they are HERGs, and all eight are extended in the optical. Of the 12 sources which are not detected in the X-ray, 11 are extended in the optical and one is point-like. 

\begin{table*}
%
% adjust \tabcolsep so table fits in page width...
%
\renewcommand{\tabcolsep}{2.0pt}
\caption{Summary of the properties of host galaxies of 10C sources.}\label{tab:obj_properties}
\smallskip
\centering
\begin{tabular}{lcccccclccccc}\cline{1-6}\cline{8-13}
\vpad
10C ID & Optical         & Lacy et al. & X-ray   & Spectral      & Overall            & \hbox{\ } & 10C ID & Optical         & Lacy et al. & X-ray  & Spectra       & Overall            \\
       & compact.$^{\rm a}$ &  area$^{\rm b}$   & detection$^{\rm c}$ & class.$^{\rm d}$ & class.$^{\rm e}$ &           &        & compact.$^{\rm a}$ & area$^{\rm b}$    & detection$^{\rm c}$ & class.$^{\rm d}$ & class.$^{\rm e}$\\\cline{1-6}\cline{8-13}
\vpad
10C J104320+585621 & E & Y & Y &L& LERG        & &  10C  J105040+573308 &   &   &   & &        \\
10C J104328+590312 & E & N & N & & LERG        & &  10C  J105042+575233 &   &   &   & &        \\
10C J104344+591503 & E & N & N & & LERG        & &  10C  J105050+580200 & E & Y &   & & HERG   \\
10C J104428+591540 &   & N & Y & & HERG        & &  10C  J105053+583233 &   & Y &   & & HERG   \\
10C J104441+591949 & E & N & N & & LERG        & &  10C  J105054+580943 &   &   &   & &        \\
10C J104451+591929 & E & N & N & & LERG        & &  10C  J105058+573356 &   &   & N & & LERG   \\
10C J104528+591328 & E & Y & Y & & HERG        & &  10C  J105104+574456 &   &   &   & &        \\
10C J104539+585730 & E & N & N & & LERG        & &  10C  J105104+575415 & P & Y &   &H& HERG   \\
10C J104551+590838 & P &   & N & & HERG        & &  10C  J105107+575752 & E & N &   & & LERG   \\
10C J104624+590447 & E &   & N & & LERG        & &  10C  J105115+573552 &   &   & Y & & HERG   \\
10C J104630+582748 & E & N &   &L& LERG        & &  10C  J105121+582648 &   &   &   & &        \\
10C J104633+585816 & E & Y & Y & & HERG        & &  10C  J105122+570854 &   &   &   & &        \\
10C J104648+590956 &   & Y & N & & HERG        & &  10C  J105122+584136 &   &   &   & &        \\
10C J104700+591903 &   &   & Y & & HERG        & &  10C  J105122+584409 & P & Y &   & & HERG   \\
10C J104710+582821 & E & Y &   & & HERG        & &  10C  J105128+570901 & E & N &   &L& LERG   \\
10C J104718+585119 & E &   & Y & & HERG        & &  10C  J105132+571114 &   & N &   &L& LERG   \\
10C J104719+582114 & P & Y &   &H& HERG        & &  10C  J105136+572944 &   & Y & Y & & HERG   \\
10C J104733+591244 &   &   & Y & & HERG        & &  10C  J105138+574957 &   &   &   & &        \\
10C J104737+592028 & E & N & N & & LERG        & &  10C  J105139+580757 &   &   &   & &        \\
10C J104741+584811 &   &   & N & & LERG        & &  10C  J105142+573447 & E & N & Y & & HERG   \\
10C J104742+585318 & E & N & N & & LERG        & &  10C  J105142+573557 & E &   & Y & & HERG   \\
10C J104751+574259 & E &   &   & &             & &  10C  J105144+573313 &   &   & N & & LERG   \\
10C J104802+574117 & E & N &   & & LERG        & &  10C  J105148+573245 & E & Y & Y & & HERG   \\
10C J104822+582436 & E & N &   & & LERG        & &  10C  J105206+574111 &   &   & Y & & HERG   \\
10C J104824+583029 & E & N &   & & LERG        & &  10C  J105215+581627 & E &   &   & &        \\
10C J104826+584838 & E &   & N & & LERG        & &  10C  J105220+585051 & E &   &   & &        \\
10C J104836+591846 & P & Y &   & & HERG        & &  10C  J105225+573323 & E & N & N &L& LERG   \\
10C J104844+582309 & E & N &   & & LERG        & &  10C  J105225+575507 &   &   &   & &        \\
10C J104849+571417 & E & N &   & & LERG        & &  10C  J105237+573058 &   &   & N & & LERG   \\
10C J104856+575528 &   &   &   & &             & &  10C  J105240+572322 & E & N & Y & & HERG   \\
10C J104857+584103 &   &   &   & &             & &  10C  J105243+574817 & E & N &   & & LERG   \\
10C J104906+571156 &   &   &   & &             & &  10C  J105327+574546 & E &   &   & &        \\
10C J104918+582801 & P & Y &   &H& HERG        & &  10C  J105341+571951 & E & N & N & & LERG   \\
10C J104927+583830 &   &   &   & &             & &  10C  J105342+574438 & E & N &   & & LERG   \\
10C J104934+570613 & E & N &   & & LERG        & &  10C  J105400+573324 &   &   & Y & & HERG   \\
10C J104939+583530 & E & Y &   &H& HERG        & &  10C  J105425+573700 & E & Y &   &H& HERG   \\
10C J104943+571739 & E & Y &   &H& HERG        & &  10C  J105437+565922 &   &   &   & &        \\
10C J104954+570456 & E & Y &   & & HERG        & &  10C  J105441+571640 &   &   &   & &        \\
10C J105000+585227 &   &   &   & &             & &  10C  J105510+574503 & E & N &   & & LERG   \\
10C J105007+572020 & E &   &   & &             & &  10C  J105515+573256 &   &   &   & &        \\
10C J105007+574251 & E &   &   & &             & &  10C  J105520+572237 &   & Y &   & & HERG   \\
10C J105009+570724 &   &   &   & &             & &  10C  J105527+571607 & E & N &   &L& LERG   \\
10C J105020+574048 & E & N &   & & LERG        & &  10C  J105535+574636 & E & Y &   & & HERG   \\
10C J105028+574522 & P & N &   &L& LERG        & &  10C  J105550+570407 & E & N &   &L& LERG   \\
10C J105034+572922 &   &   &   & &             & &  10C  J105604+570934 &   &   &   & &        \\
10C J105039+572339 & P & Y &   &H& HERG        & &  10C  J105627+574221 & P & Y &   & & HERG   \\
10C J105039+574200 & E &   &   & &             & &  10C  J105653+580342 & E & N &   &L& LERG   \\
10C J105039+585118 & E & N &   &L& LERG        & &  10C  J105716+572314 &   &   &   & &        \\\cline{1-6}\cline{8-13}
\end{tabular}
\vpad
\begin{flushleft}
 Notes:\\
 a) E = source is classified as extended, P = source is classified as point-like (see Section~\ref{section:opt-compact}).
 b) Y = source is located inside the \citeauthor{2004ApJS..154..166L} AGN area on the mid-infrared colour--colour
 diagram, N = source is located outside this region (including sources with limits which must lie
 outside this region), see Section~\ref{section:colour-colour}.
 c) Y = source is detected by an X-ray survey, N = source lies inside X-ray survey area but is not
 detected (see Section~\ref{section:x-ray}).
 d) H = source is classified as a HERG based on its optical spectrum, L = source is classified as a LERG based on its optical spectrum. (See Section~\ref{section:spectra}).
 e) HERG = source is classified as a `probable HERG', LERG = source is classified as a `probable LERG' (see Section~\ref{section:HERG_overall_class}).
\end{flushleft}
\end{table*}

\begin{figure}
\centerline{\includegraphics[width=\columnwidth]{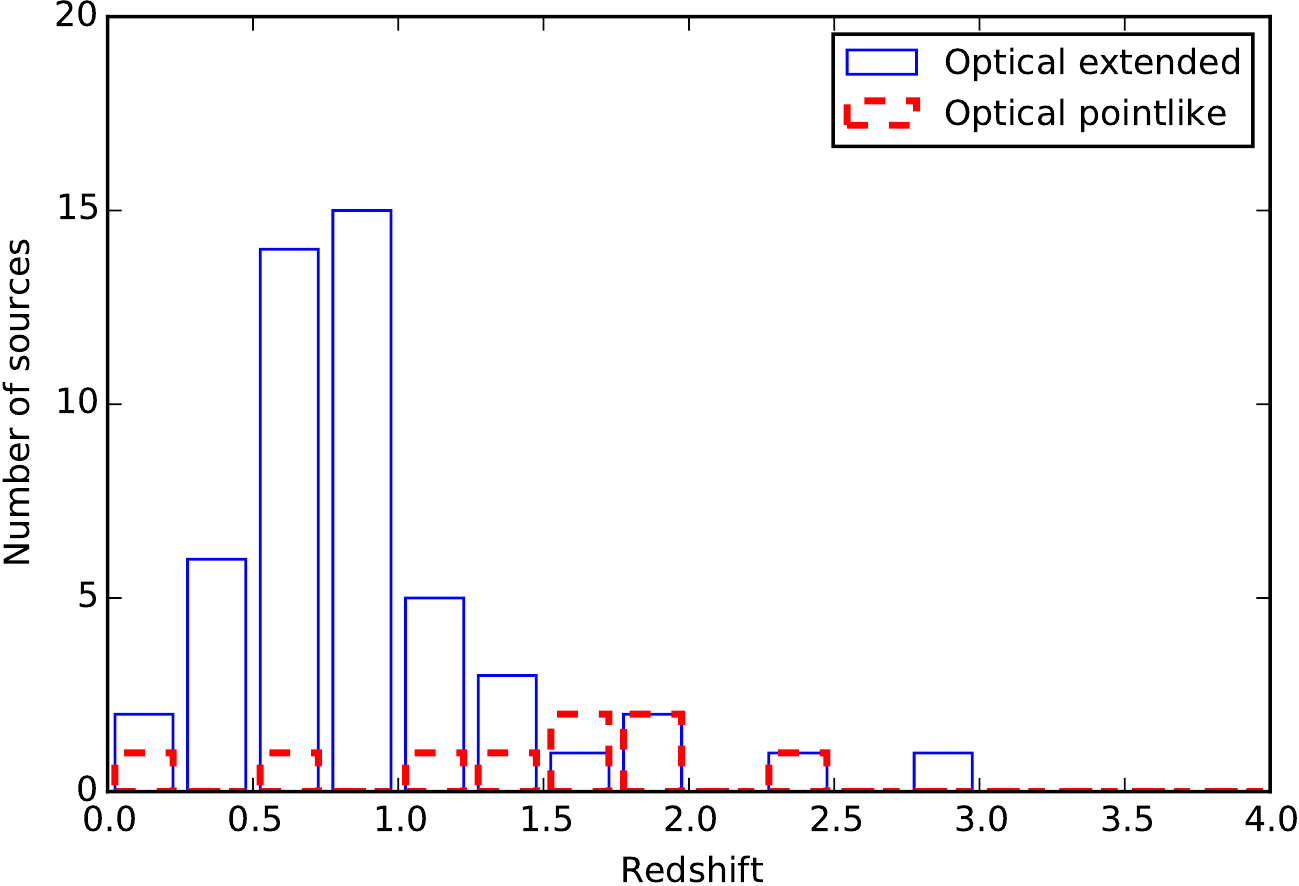}}
\caption{Redshift distribution for sources which are point-like and extended in the optical. Note that the 37 sources which do not have an optical counterpart and/or redshift information are not included in this figure.}\label{fig:opt_z}
\end{figure}

\begin{figure*}
\centerline{\includegraphics[width=5.50cm]{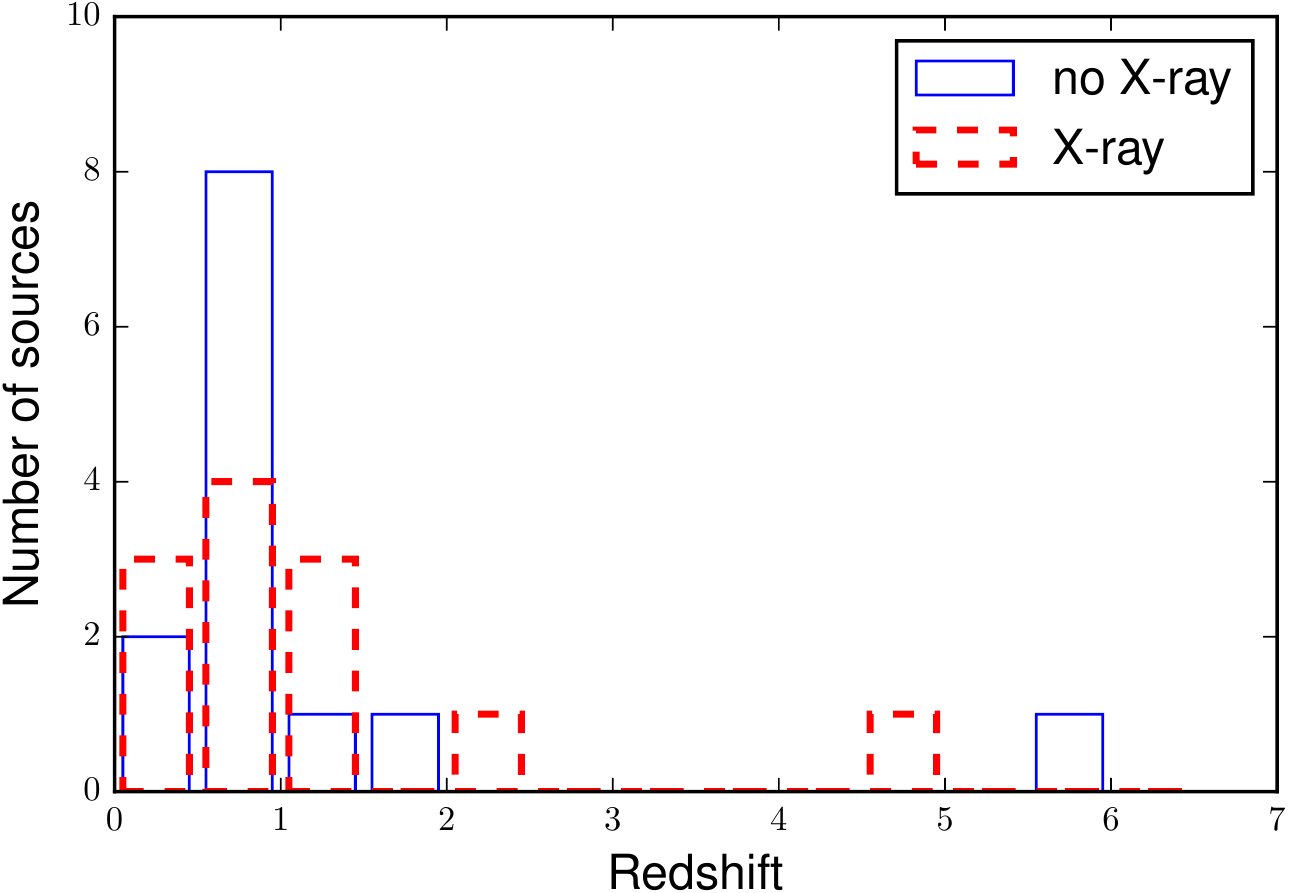}
         \quad
         \includegraphics[width=5.50cm]{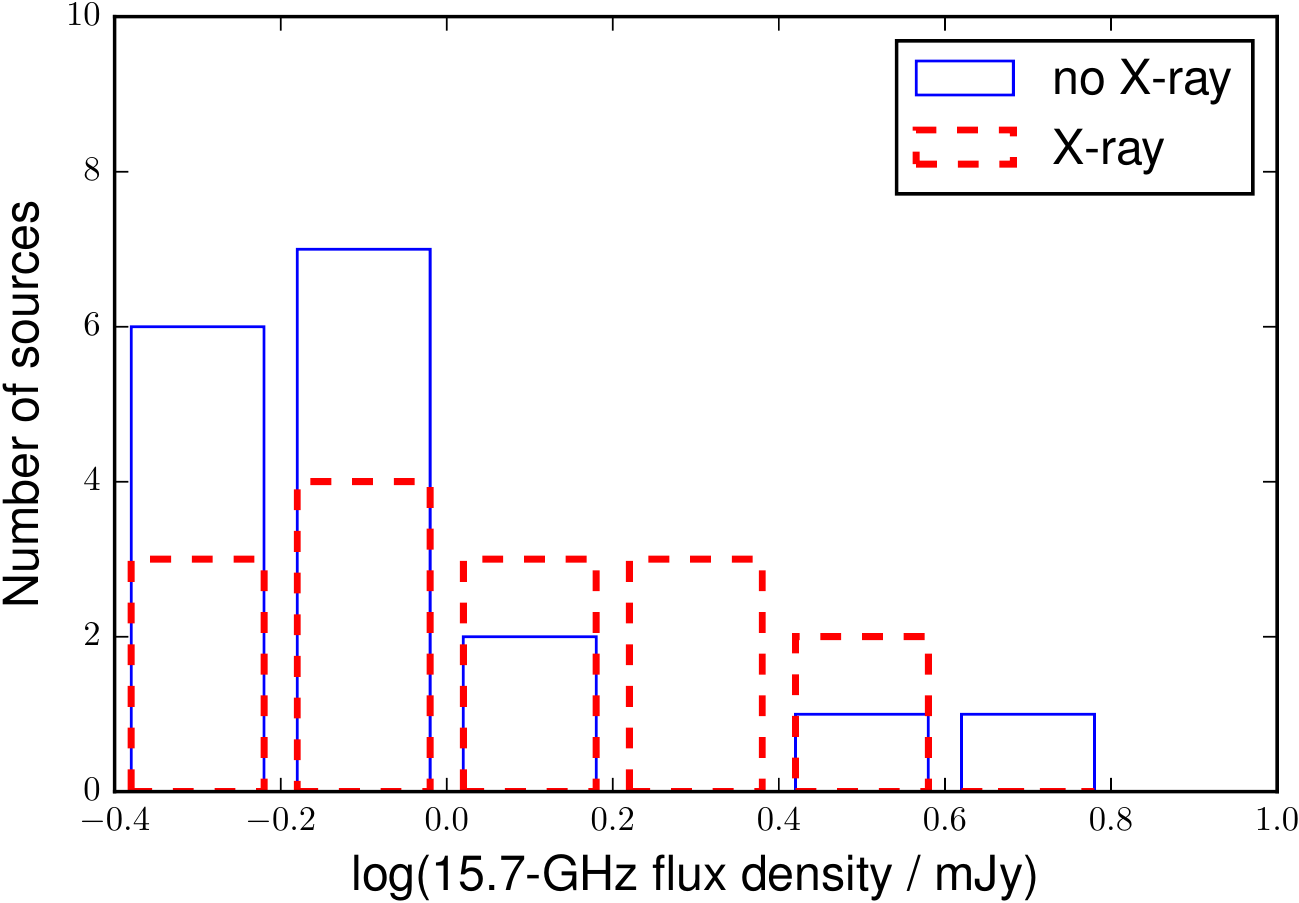}
         \quad
         \includegraphics[width=5.50cm]{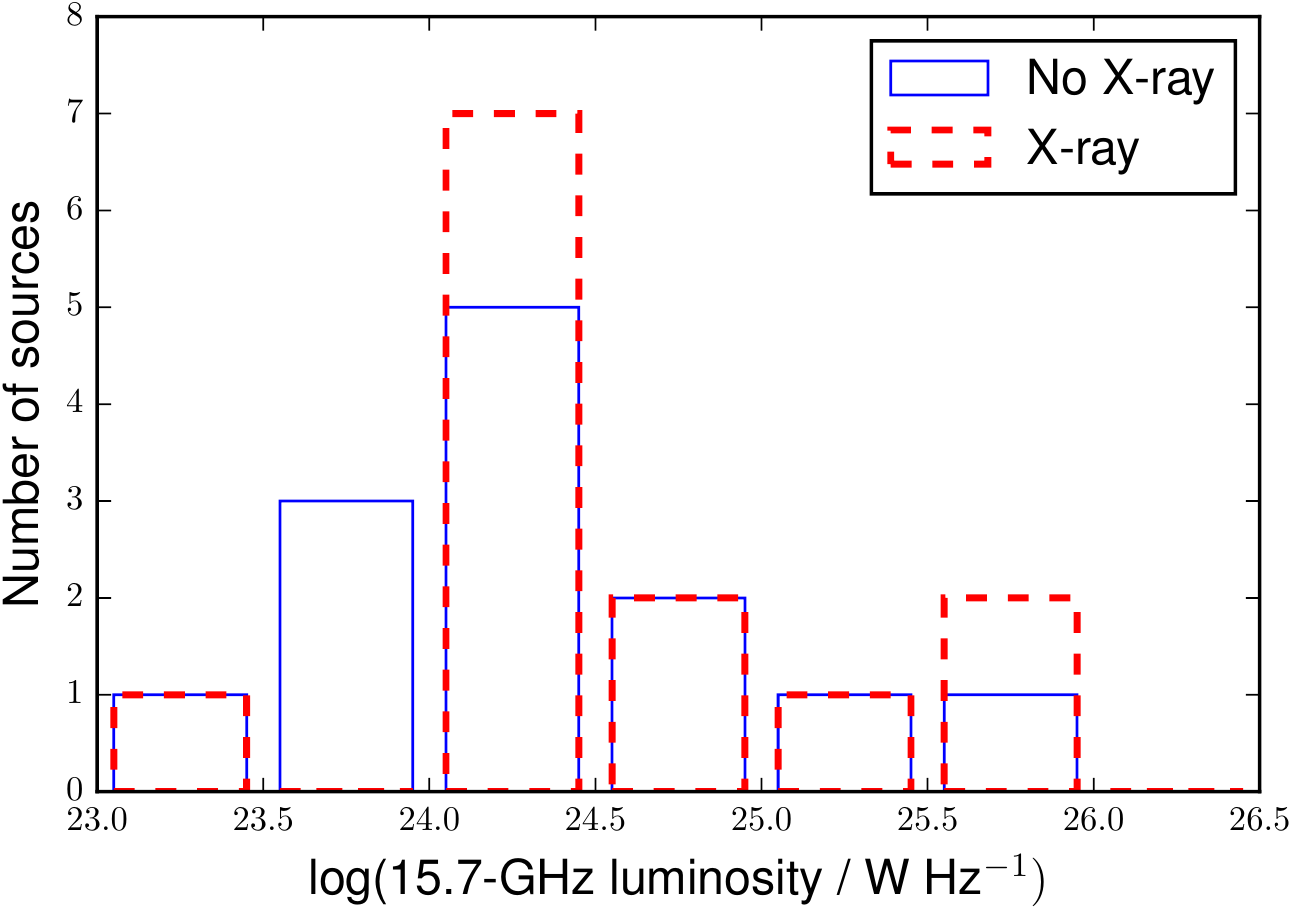}}
\caption{Redshift, 15.7-GHz flux density and 15.7-GHz luminosity distributions for sources with and without X-ray detections. For flux density, the 32 sources which lie inside either of the two X-ray survey areas are shown, while for redshift and luminosity only the 26 sources which are inside the X-ray survey areas and have redshift values available are included.}\label{fig:xray_z_S}
\end{figure*}

\begin{figure*}
\centerline{\includegraphics[width=5.50cm]{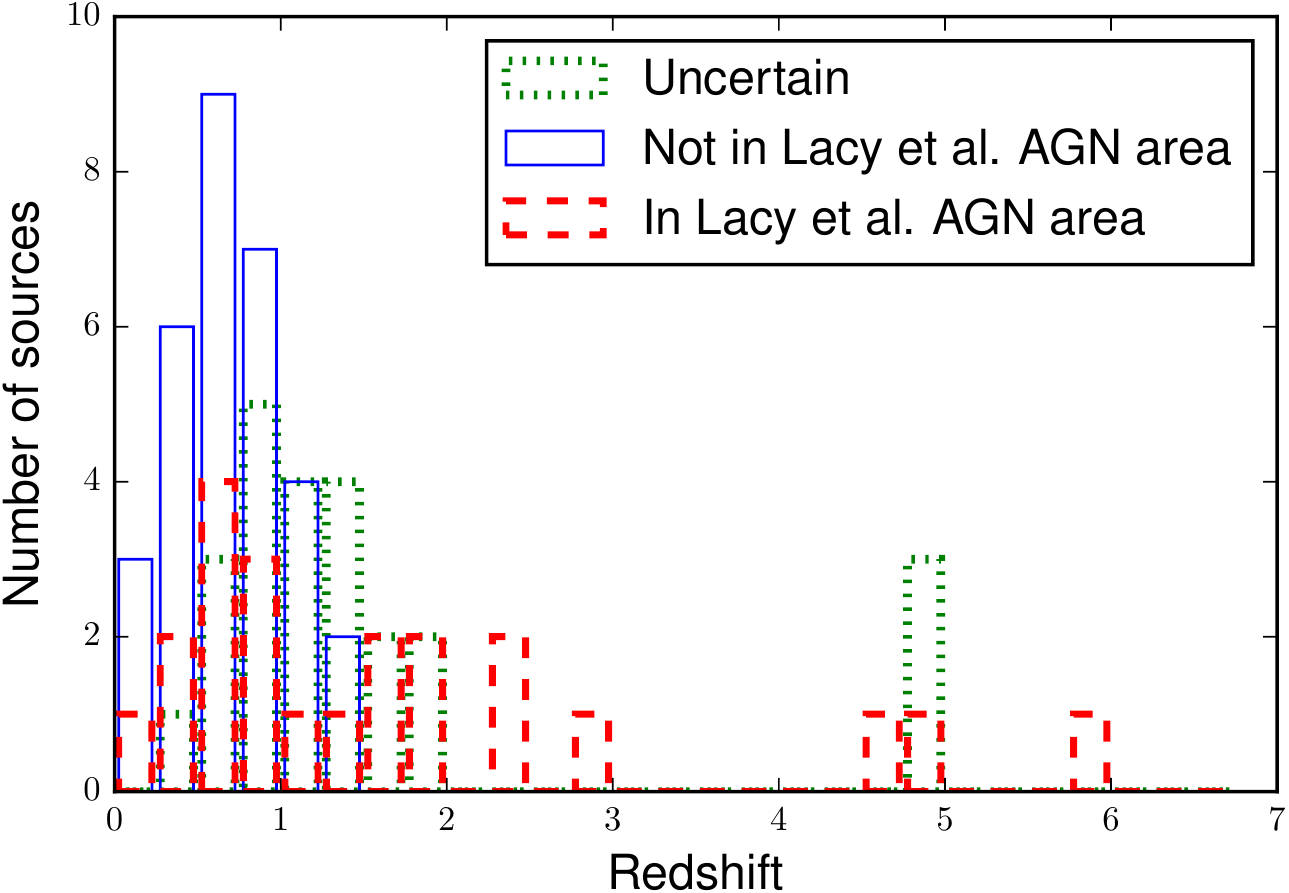}
         \quad
         \includegraphics[width=5.50cm]{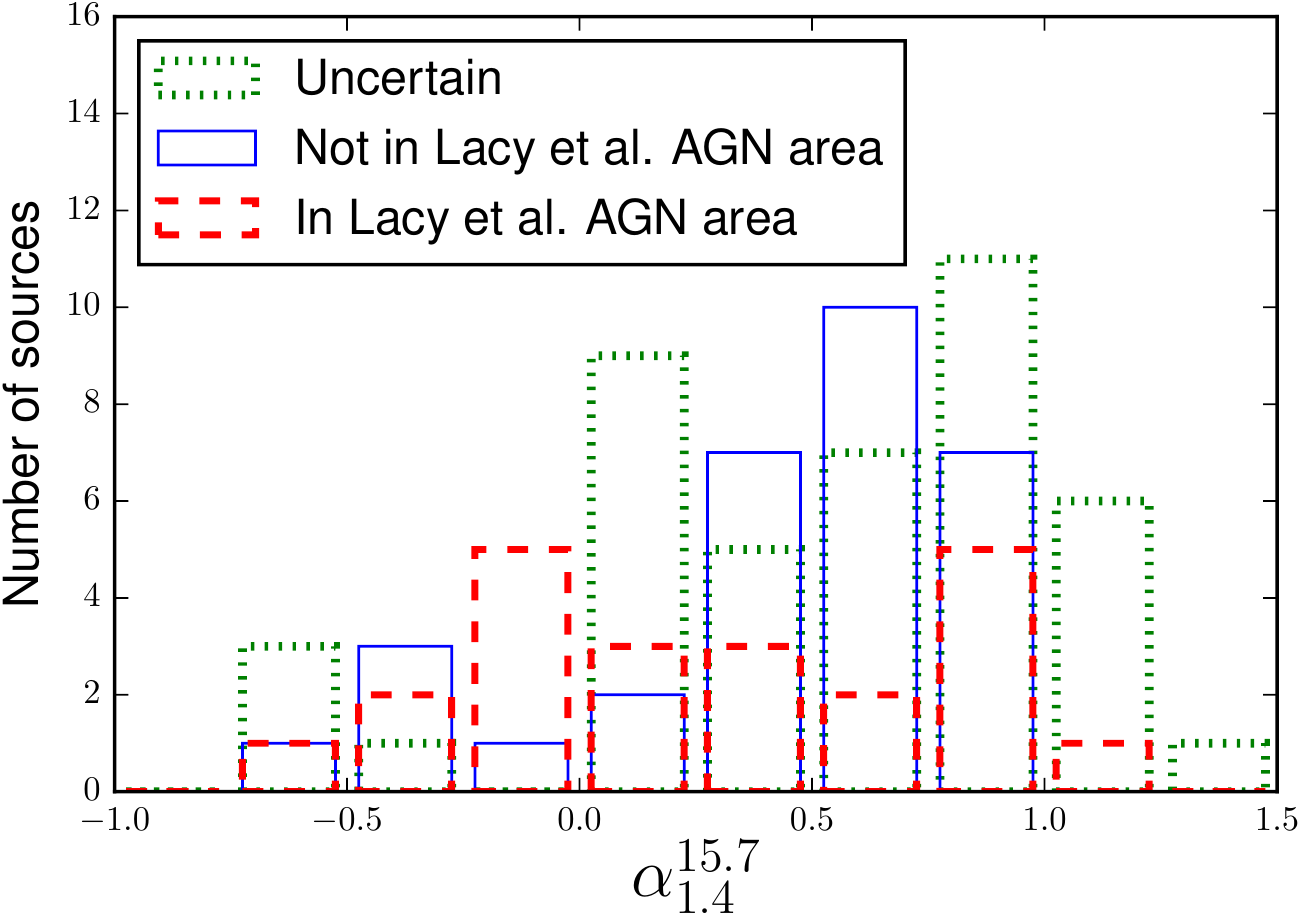}
         \quad
         \includegraphics[width=5.50cm]{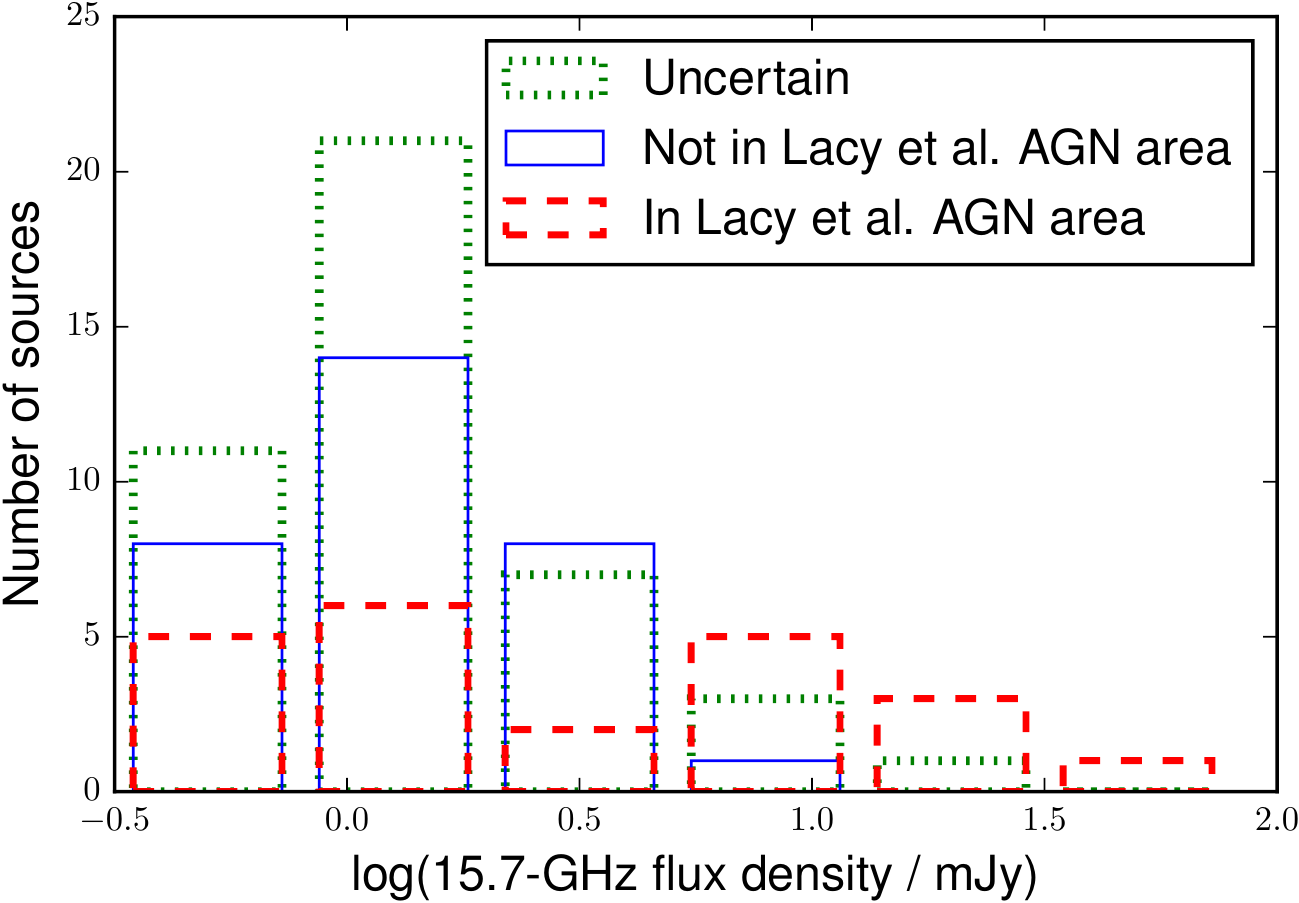}}
\caption{Redshift, spectral index and 15.7-GHz flux density distributions for different object types classified according to their positions on the Lacy et al. mid-infrared colour--colour diagram; the `uncertain' classification refers to sources whose position on the diagram is uncertain as they are undetected in one or more mid-infrared bands. All 96 sources are included in the flux density and spectral index plots, while the 77 sources with redshift values available are included in the redshift plot.}\label{fig:lacy_z}
\end{figure*}

\subsection{Comparison with spectra}\label{section:spectra}

Spectra are available from the Sloan Digital Sky Survey Data Release 12 (SDSS DR 12; \citealt{2015ApJS..219...12A}) for 17 out of the 96 sources in this sample, which provides a more direct way to classify these sources as HERGs or LERGs. Following \citet{1994ASPC...54..201L,1997MNRAS.286..241J} and \citet{2016MNRAS.457..730P} we make use of the 5007~$\AA$ [OIII] line to distinguish between the two populations; sources with [OIII] equivalent widths greater than $5~\AA$ are classified as HERGs, while sources without measurable [OIII] widths or with widths less than this value are classified as LERGs.

Out of the 17 sources classified this way, 10 are HERGs and seven are LERGs. In 16 out of 17 cases (94 per cent), these spectroscopic classifications agree with the classifications obtained using our other methods, demonstrating that they provide a reliable way of distinguishing between HERGs and LERGs. In one case, however, there is some discrepancy between the classifications; source 10C J104320+585621 has an [OIII] line width of 1.25~$\AA$ so is classified as a LERG from its spectrum, but lies inside the \citeauthor{2004ApJS..154..166L} AGN area on the mid-infrared colour--colour diagram and is detected in the X-ray, suggesting it is a HERG. The spectroscopic classification for this source is used in the overall classifications discussed in the following section.

\subsection{Overall classifications}\label{section:HERG_overall_class}

The classification methods discussed in this section can be combined to produce an overall classification of radio galaxy type. Sources which do not have any of the three pieces of information discussed in Sections~\ref{section:opt-compact}, \ref{section:colour-colour}, \ref{section:x-ray} or a spectroscopic classification available are not classified. For the 17 sources with a spectroscopic classification available, this classification is used. For the remaining sources, any source which displays any of the three indicators of high-excitation behaviour (i.e.\ is point-like in the optical, lies inside the \citeauthor{2004ApJS..154..166L} AGN area, or is detected in the X-ray observations) is classified as a `probable HERG'. Any source which has mid-infrared information available and lies outside the \citeauthor{2004ApJS..154..166L} AGN area, or is inside the X-ray survey area but is not detected is classified as a `probable LERG'. Sources which only have optical compactness information available and are extended are not classified as they could be either HERGs or LERGs. This results in 32 sources being classified as probable HERGs, 35 as probable LERGs, and 29 sources are not classified at this stage. The details are given in Table~\ref{tab:obj_properties} and selected properties of the sources are shown in Table~\ref{tab:properties}.

%----------------------------------------------------------------------------------------------------%
\section{Properties of HERGs and LERGs}\label{section:object_properties}

\subsection{Properties of HERGs and LERGs split using the three different methods}\label{section:HERG_3methods}

Fig.~\ref{fig:opt_z} shows the redshift distribution of objects split according to their optical compactness. Although the numbers are small it can be seen that there is a higher proportion of the high-redshift sources which are point-like ($7/20$, 35 per cent, of sources with $z > 1$, compared to $2/37$, 5 per cent, of sources with $z < 1$). This can be attributed to the fact that we expect the objects which appear point-like in the optical to be quasars; as these sources are rare but very luminous they tend to be found preferentially at higher redshifts where larger volumes are being sampled. There seems to be no significant trend in spectral index with optical compactness, although of the 11 sources with steeply falling spectra ($\alpha^{15.7}_{1.4} > 0.8$) or the four with sharply rising ($\alpha < -0.4$) spectra, none are point-like. Both the point-like and the extended objects cover the same range of 15.7-GHz flux density and 1.4-GHz luminosity.

Fig.~\ref{fig:xray_z_S} shows the properties of the 32 objects inside either of the two X-ray survey areas which are and are not detected in the X-ray observations (note that the two different X-ray surveys are combined here). A larger proportion of the higher-redshift sources are detected in the X-ray observations, with $6/9$ (67 per cent) sources with $z>1$ detected, compared to $7/17$ (41 per cent) sources with $z < 1$. This could be because some of the high-redshift sources are quasars and therefore do not suffer any absorption. There is no clear difference in the spectral indices of those sources which are detected in the X-ray and those which are not. The 15.7-GHz flux densities, however, tend to be higher for the X-ray detected sources, with $8/12$ (67 per cent) sources with $S_{15.7~\rm GHz} > 1~\rm mJy$ detected, compared with $7/20$ (35 per cent) sources with $S_{15.7~\rm GHz} < 1~\rm mJy$. Having said this, some of the faintest sources are X-ray detected, with three out of five sources in the faintest flux density bin with $S_{15.7~\rm GHz} < 0.63~\rm mJy$ detected in the X-ray observations. There is also some indication that fewer X-ray-detected sources have low 15.7-GHz luminosities. The small number of sources (32) in the X-ray sample must be borne in mind when considering the significance of these results.

Fig.~\ref{fig:lacy_z} shows the properties of sources which lie inside and outside the \citeauthor{2004ApJS..154..166L} AGN area on the mid-infrared colour--colour diagram (see Section~\ref{section:colour-colour}). There is a clear trend with redshift visible, with all 17 sources with $z > 1.5$ lying inside the \citeauthor{2004ApJS..154..166L} AGN area or having uncertain positions. There is also a trend with spectral index, with a larger proportion of the flat spectrum sources lying inside the \citeauthor{2004ApJS..154..166L} AGN area ($14/41$, 34 percent, of flat spectrum sources lie inside the \citeauthor{2004ApJS..154..166L} AGN area, compared to $8/38$, 21 percent, of the steep spectrum sources). The 15.7-GHz flux densities of the samples are also significantly different, with the Lacy AGN tending to have higher flux densities. This suggests that the composition of the population may be changing with 15.7-GHz flux density, with a larger proportion of sources with larger flux densities classified as HERGs. There is also a higher proportion of HERGs with larger 1.4-GHz luminosities, which is unsurprising given their higher flux densities and redshifts.

All three methods for identifying object types discussed here provide useful information about the population. The \citeauthor{2004ApJS..154..166L} colour--colour diagram appears to be the most effective method of splitting objects with different radio properties. Sources located inside the \citeauthor{2004ApJS..154..166L} AGN area tend to have flatter spectra ($\alpha^{15.7}_{1.4}$), be found at higher redshifts, and have higher 15.7-GHz flux densities. These results will be discussed in more detail in the next section after the properties of the overall classifications are presented.

\subsection{Properties by overall classification}\label{section:HERG_overall}

\begin{figure*}
\centerline{\includegraphics[width=8cm]{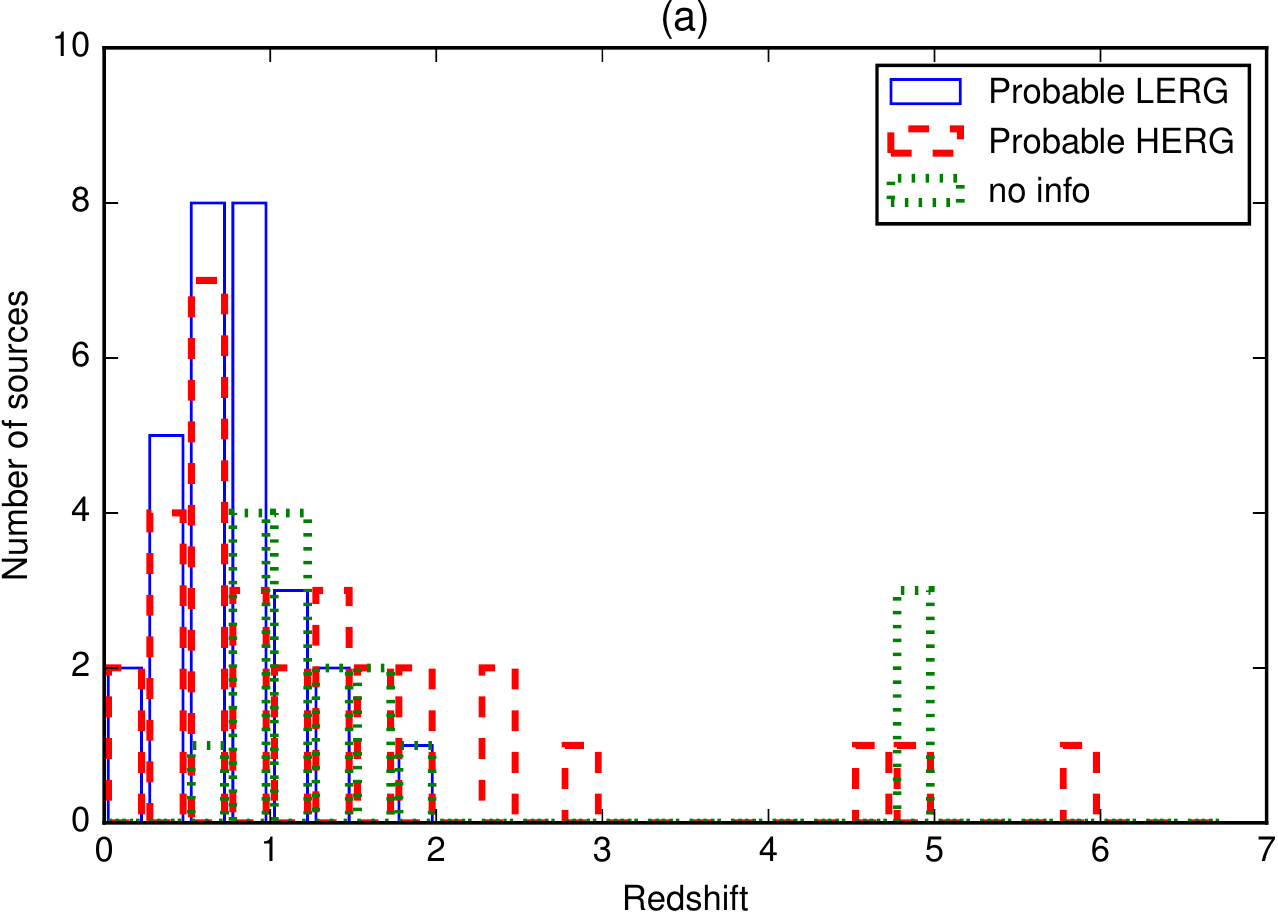}
            \quad
            \includegraphics[width=8cm]{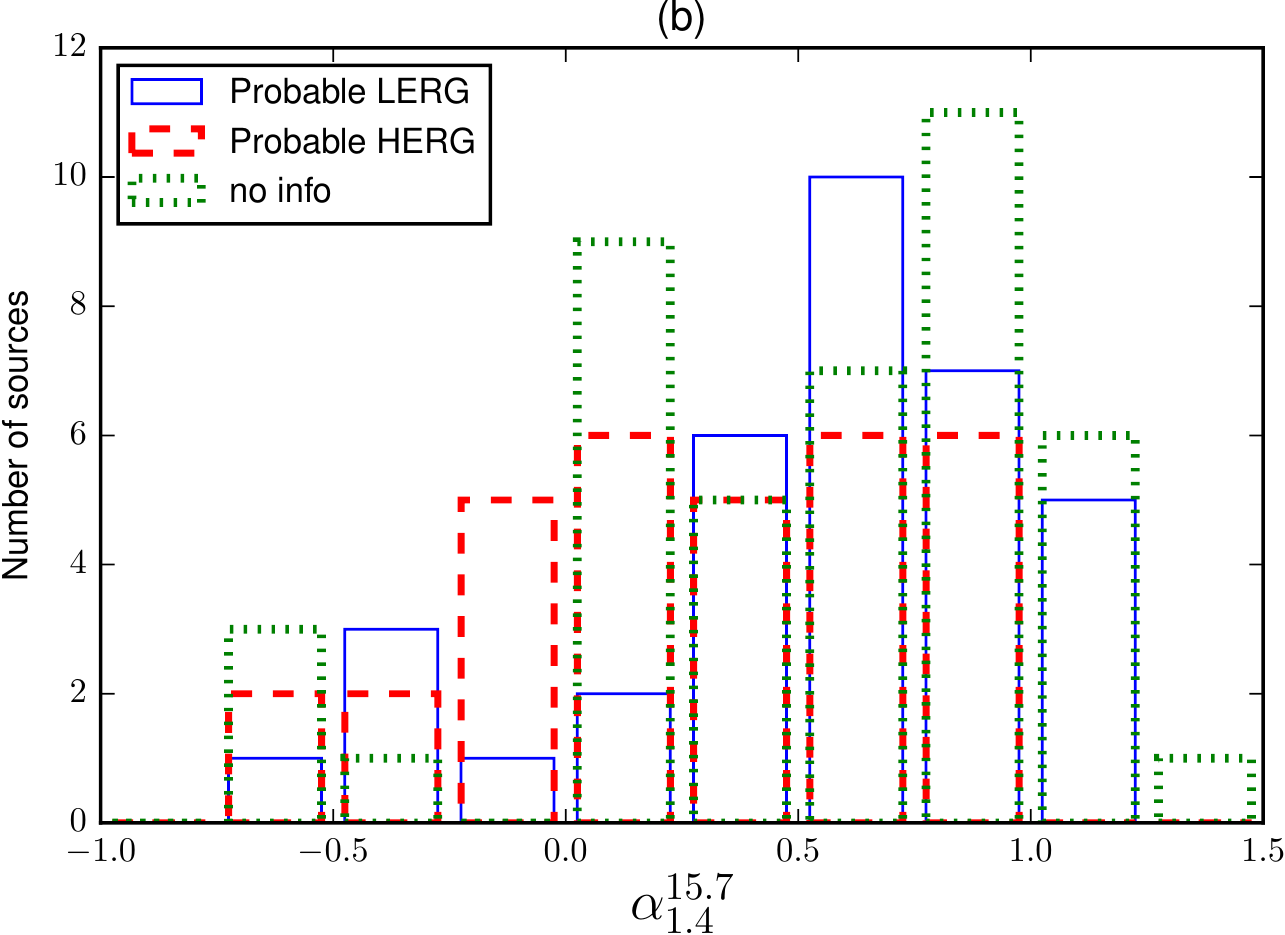}}
\smallskip
\centerline{\includegraphics[width=8cm]{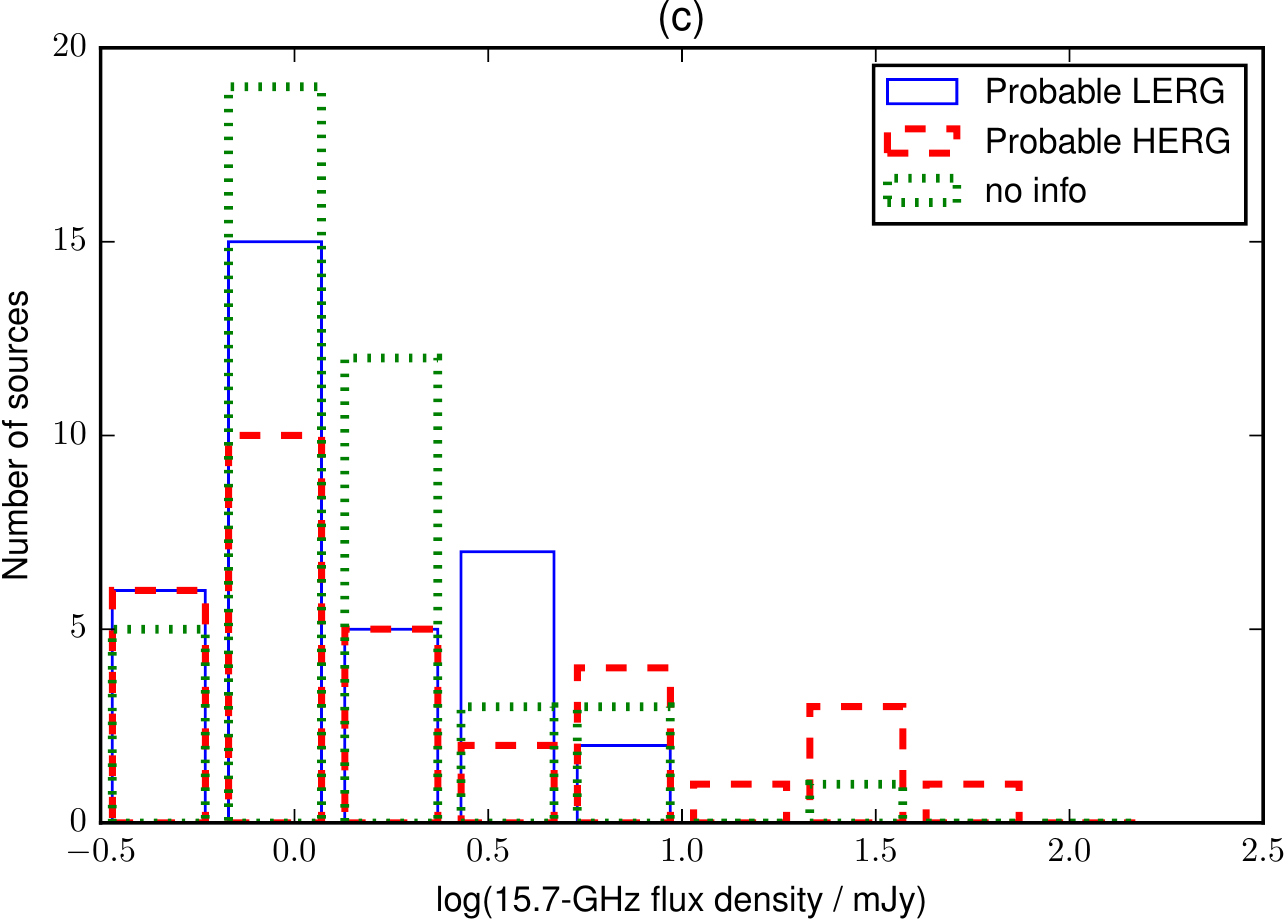}
            \quad
            \includegraphics[width=8cm]{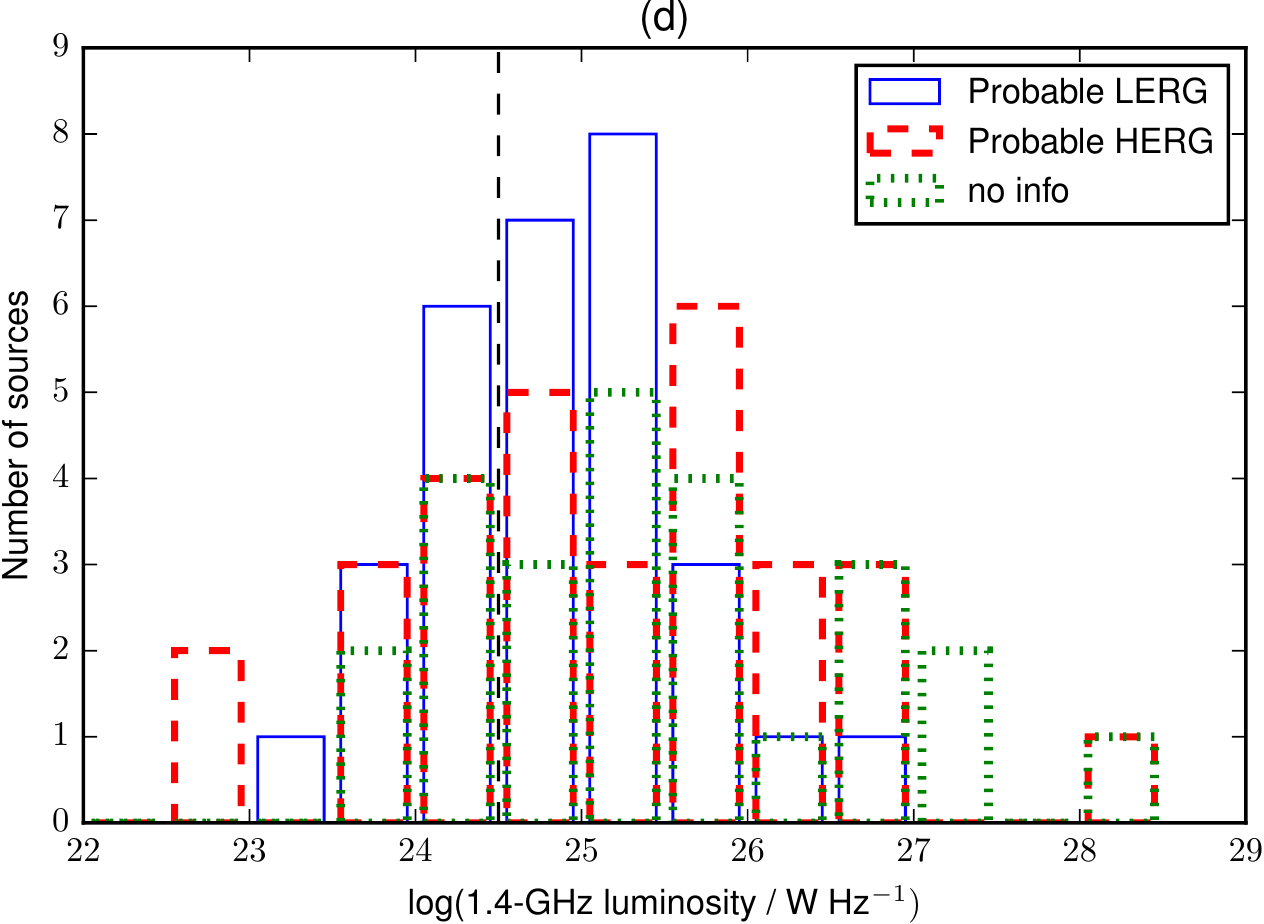}}
\caption{Radio properties sources classified as HERGs and LERGs, and those with insufficient information available to be classified. The top left panel shows the redshift distributions and the top right panel shows the spectral index distributions. The bottom left panel shows the 15.7-GHz flux density distributions and the bottom right panel shows the 1.4-GHz luminosity distributions. The dashed line in the bottom right panel is at ${\rm log}(L_{1.4~\rm GHz} / \rm W \, Hz^{-1}) = 24.5$, the dividing luminosity between FRI and FRII sources. {All 96 sources are included in the flux density and spectral index plots, while the 77 sources with redshift information available are included in the redshift and luminosity plots.} }\label{fig:HERG_properties}
\end{figure*}

Selected properties of the sample of 96 sources studied in this paper are shown in Table~\ref{tab:properties}. The properties of the probable HERGs and probable LERGs are shown in Fig.~\ref{fig:HERG_properties}. The 29 sources which could not be classified as HERGs or LERGs as there is not enough information available are also shown in this figure. Panel (a) of Fig.~\ref{fig:HERG_properties} shows the redshift distribution of the HERGs and LERGs; the HERG distribution has a high redshift tail, whilst there are no LERGs with $z > 2$. A Kolmogorov--Smirnov (K--S) test on these two distributions shows that there is only a 2 per cent chance of the null hypothesis. Panel (b) shows the spectral index ($\alpha^{15.7}_{1.4}$) distributions, which indicates that the HERGs tend to have flatter spectra between 1.4 and 15.7~GHz than the LERGs. A K--S test on these two spectral index distributions shows that there is only a 1 per cent chance that they are drawn from the same population. The 15.7-GHz flux density distribution is shown in panel (c); the distributions peak at similar values but the HERG distribution extends to higher flux densities than the LERG distribution (no LERGs have a 15.7~GHz flux density greater than 6~mJy). The 1.4-GHz luminosity distributions are shown in panel (d), and are similar for HERGs and LERGs. Although the HERGs have higher redshifts and 15.7-GHz flux densities than the LERGs, they have flatter spectra -- these combine to produce similar 1.4-GHz luminosity distributions for the two groups. Both HERGs and LERGs are found on either side of the FRI/FRII dividing luminosity (shown by the vertical dashed line).

\citet{2012MNRAS.421.1569B} found that although both HERGs and LERGs display the full range of luminosities, HERGs dominate at luminosities greater than $L_{1.4~\rm GHz} \approx 10^{26}~\rm W \, Hz^{-1}$, while LERGs dominate at lower luminosities. Their study, however, focused on nearby radio galaxies (median redshift of 0.16) so may not be directly comparable to the 10C sample, which has a median redshift of 0.91 (Paper II). \citet{2015MNRAS.447.1184F} studied a sample of 27 HERGs and LERGs at $z \sim 1$. They found that HERGs and LERGs have very similar luminosity distributions up to $L_{151~\rm{MHz}} \sim 4 \times 10^{28} \, \rm W \, Hz^{-1}$ (note that this is at 151~MHz so is equivalent to $\sim 5 \times 10^{27} \, \rm W \, Hz^{-1}$ at 1.4~GHz), after which HERGs dominate. This is consistent with the findings here, where the luminosity distributions of the two samples are similar, and out of the two sources with $L_{1.4~\rm GHz} \gtrsim 5 \times 10^{27} \, \rm W \, Hz^{-1}$, one is a HERG and one is not classified.

\begin{figure}
\centerline{\includegraphics[width=\columnwidth]{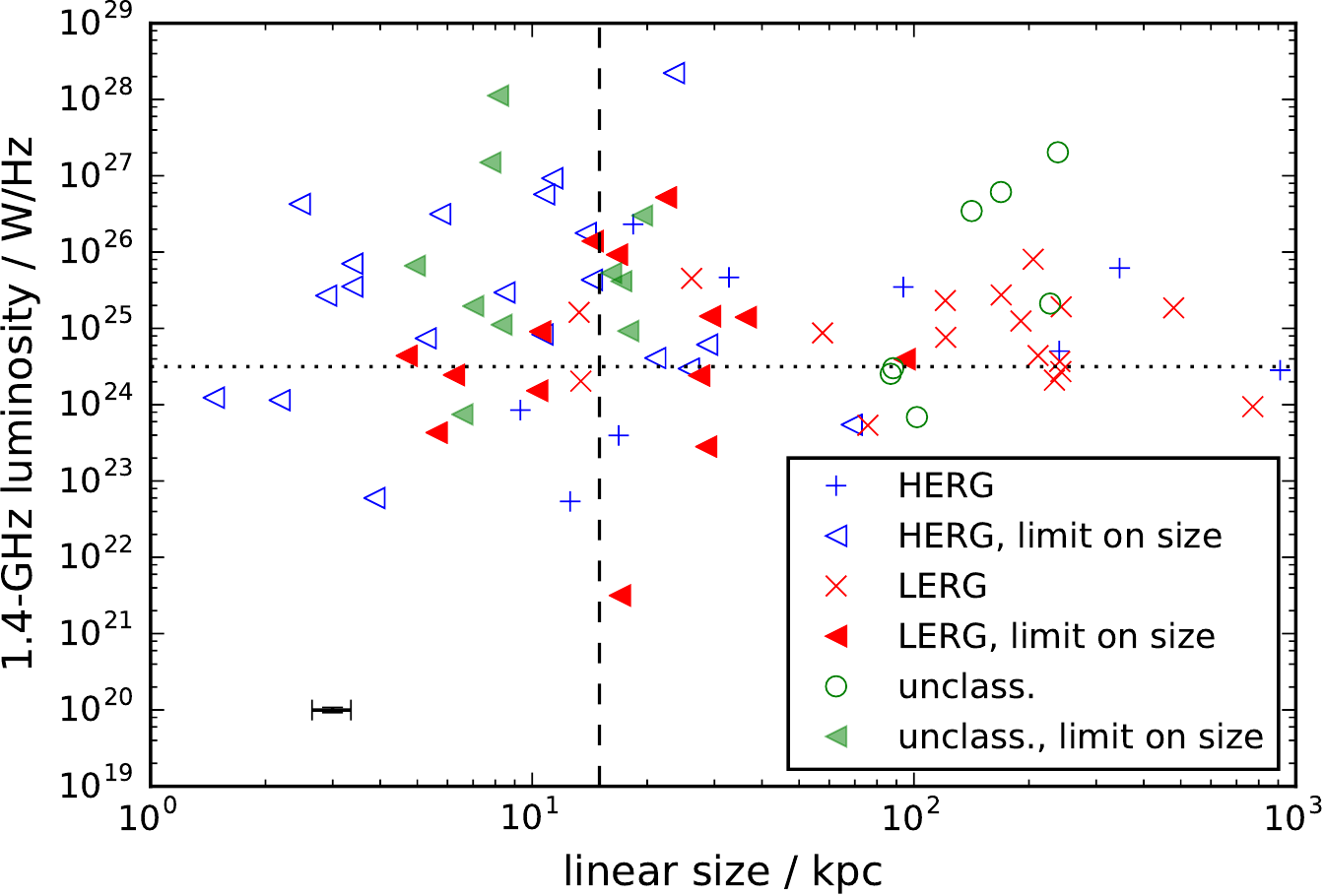}}
\caption{1.4-GHz luminosity as a function of linear size with HERGs, LERGs and unclassified sources shown separately. Sources with upper limits on their linear size are shown as triangles, and could move to the left. The vertical dashed line is at $d = 15$~kpc; CSS sources are typically smaller than this value. The horizontal dotted line is at ${\rm log(}L_{1.4~\rm GHz}  / \rm W \, Hz^{-1}) = 24.5$, the typical luminosity divide between FRI and FRII sources. For clarity the individual error bars have been omitted but a point representing the median uncertainties in luminosity and linear size is included in the bottom left corner. The 77 sources with redshift information are shown in the plot. }\label{fig:lum14_linsize_HERG}
\end{figure}

The linear sizes, which were calculated in Section 6.4 in Paper II, are plotted against luminosity in Fig.~\ref{fig:lum14_linsize_HERG} with different symbols for HERGs and LERGs. Although both HERGs and LERGs cover the full range of linear sizes, the HERGs tend to be smaller, with 53 per cent of HERGs having $d<15~\rm kpc$ compared to only 23 per cent of LERGs (although as some of the linear sizes are upper limits these percentages could be higher in both cases). There is some indication that the HERGs and LERGs lie in different regions on this diagram. The top left quadrant (marked by the dotted and dashed lines), where small and luminous sources with $d < 15~\rm kpc$ and $L_{1.4~\rm GHz} > 10^{24.5} \rm W \, Hz^{-1}$ are found, is dominated by HERGs, which comprise 12 out of the 16 sources, while in the bottom right quadrant, where large and less luminous sources are found, four out of the six sources are LERGs (excluding the sources with upper limits on their sizes).

In summary, the HERGs in the 10C sample tend to be at larger redshifts, and have larger 15.7-GHz flux densities, flatter spectra and smaller linear sizes.  The tendency for HERGs to have flat spectra, coupled with their smaller linear sizes, suggests that they are dominated by emission from their cores. However, the relatively large proportion of unclassified sources (29 out of 96 sources, these are shown by green dashes in Fig.~\ref{fig:HERG_properties}), must be taken into account when considering the significance of these results. If, for example, these unclassified sources turned out to all be HERGs this would significantly alter these conclusion. The morphology of these sources is discussed in more detail in the next section. 

%------------------------------------------------------------------------------%
\section{Classifications by radio morphology}\label{section:FR-sources}

\subsection{Classifying the sources}

\begin{table}
\caption{Spectral index classifications used in this paper.}\label{tab:spec_class}
\medskip
\centering
\begin{tabular}{llr}\hline
Spectrum & Class & No. (per cent)\\\hline
\vpad
$\alpha^{15.7}_{1.4} > 0$ and $\alpha^{1.4}_{0.61} > 0^\ast$ & Steep (S) & 51 (53) \\
$\alpha^{15.7}_{1.4} < 0$ and $\alpha^{1.4}_{0.61} < 0^\ast$ & Rising (R) & 2 (2)  \\
$\alpha^{15.7}_{1.4} > 0$ and $\alpha^{1.4}_{0.61} < 0^\ast$ & Peak (P) & 15 (16)  \\
$\alpha^{15.7}_{1.4} < 0$ and $\alpha^{1.4}_{0.61} > 0^\ast$ & Upturn (U) & 10 (10)\\
$-0.5 < \alpha^{15.7}_{1.4} < 0.5$ and  & Flat (F) & 18 (19) \\
\quad $-0.5 < \alpha^{1.4}_{0.61} < 0.5$ & &\vpad\\
\hline
\end{tabular} 
\vpad

$^\ast$ Must also not fall into `flat' category.
\end{table}

\begin{table*}
\caption{Morphological classifications used in this paper (adapted from \citealt{2014MNRAS.438..796S}).}\label{tab:morph_class}
\medskip
\centering
\begin{tabular}{llrr}\hline
Class & Criteria & Full sample & $S_{15.7~\rm GHz} < 1$~mJy\\
      &          & No. (per cent) & No. (per cent)\\\hline
\vpad
FRI & Extended emission, with peak closer to middle than edge. & 5 (5) & 0 (0)\\
FRII & Extended emission with two lobes clearly visible and distance  & 8 (8) & 2 (6)\\
     & between hotspots greater than half the total extent of the source. & \\
FRI/II (uncertain) & Two components but insufficient evidence to classify. & 5 (5) & 1 (3)\\
FR0/I (uncertain) & Extended, but no evidence for two components.   & 13 (14) & 4 (11)\\
FR0 (all) & No evidence for extended emission on arsec-scales. & 65 (68) & 29 (81)\\
\quad FR0c (candidate CSS source) & FR0 source, $\alpha^{15.7}_{1.4} > 0.5$ and not known to have $d > 15$~kpc. & 13 (14) & 3 (8)\\
\quad FR0g (candidate GPS source) & FR0 source, $\alpha^{15.7}_{1.4} < 0$ and not known to have $d > 1$~kpc. & 10 (10) & 3 (8)\\
\quad FR0u (unclassified) & All other FR0 sources. & 42 (44) & 23 (64)\\
\hline
Total & & 96 & 36\\
\hline
\end{tabular} 
\end{table*}

Radio spectral indices are a useful tool when investigating the morphology of radio sources as they provide information about the relative contributions made by the core and the lobes to the total flux density of the source. Core-dominated sources have flat spectra, while lobe-dominated sources have steeper spectra. To enable direct comparison with the results from the AT20G-6dFGS study of nearby galaxies by \citet{2014MNRAS.438..796S} we have adopted spectral classifications which are similar to theirs, as summarised in Table~\ref{tab:spec_class}. 

To split the sources in our sample into different morphological classes, GMRT 610-MHz images with a resolution of $5\times 6$~arcsec of all 96 sources were examined by eye. The sources were classified using the criteria listed in Table~\ref{tab:morph_class}. In addition to the FRI, FRII and FR0 (compact) classifications we include two `uncertain' classes: the FRI/II class, which contains sources which display two components but there is insufficient evidence to classify them as either FRI or FRII sources, and the FR0/I class, for sources which are extended but do not display evidence of two components. Linear sizes and spectral indices were then used to further divide the compact FR0 category in a similar way to \citet{2014MNRAS.438..796S}. Sources with $\alpha^{15.7}_{1.4} > 0.5$ (which therefore probably have radio spectra which peak below 1~GHz) and not known to have linear sizes $>15$~kpc are classified as candidate CSS sources (FR0c). Note that there are sources with upper limits on their linear size which are greater than 15~kpc included in this category. Sources with $\alpha^{15.7}_{1.4} < 0$ (i.e.\ likely to have a spectrum which peaks about 1~GHz) and not known to have linear sizes $>1$~kpc are classified as candidate GPS sources (FR0g); note that almost all the sources in this class have upper limits on their linear size. The remaining FR0 sources fall into the `unclassified' class (FR0u). This class includes all FR0 sources with $0< \alpha^{15.7}_{1.4} < 0.5$ and those with linear size values $>1$~kpc (and $\alpha^{15.7}_{1.4} < 0$) or $>15$~kpc (and $\alpha^{15.7}_{1.4} > 0.5$. The morphological classification of each source are listed in Table~\ref{tab:properties}, along with a selection of other source properties. 

The number of sources falling into each morphological class is summarised in Table~\ref{tab:morph_class}. 18 sources are FRI or II radio galaxies, 13 show some extended emission and the remaining 65 are FR0 sources, showing that the majority of the 10C sample are compact radio galaxies. In Paper I we found that sources with $S_{15.7~\rm GHz} < 1$~mJy have significantly flatter spectra than the brighter sources in the 10C sample, this faint subset is therefore listed separately in Table~\ref{tab:morph_class}. This shows that over 80 per cent of the sources with  $S_{15.7~\rm GHz} < 1$~mJy are compact FR0 sources, with only a small number of extended sources. This reduction in the number of extended sources at low flux densities is consistent with the fact that this population display flatter radio spectra, as extended radio structures produce steep-spectrum optically-thin synchrotron emission.

\begin{table*}
\caption{Table summarising selected properties of the 96 sources studied in this paper.}\label{tab:properties}
\small\rm
\medskip
\centering
\renewcommand{\tabcolsep}{1.6mm}
\begin{tabular}{lddddccdcdcrrccc@{}}\hline
\vpad
ID & \dhead{$S_{15.7~\rm GHz}$} & \dhead{$\sigma\_S_{15.7~\rm GHz}$} & \dhead{$\alpha^{15.7}_{1.4}$} & z & $z$            & \multicolumn{2}{c}{ang. size} & \multicolumn{2}{c}{lin. size} & \dhead{$L_{1.4~\rm GHz}$} & HERG & spec & VLBI & morph.  \\
   & \dhead{(mJy)}   & \dhead{(mJy)}              &                               &   & flag & \multicolumn{2}{c}{(arcsec)}    & \multicolumn{2}{c}{(kpc)}       & \dhead{W / Hz}          & class    & type &      & class  \\
 (1) & (2) & (3) & (4) & (5) & \multicolumn{2}{c}{(6)} & \multicolumn{2}{c}{(7)} & (8) & (9) & (10) & (11) & (12) & (13) \\     
\hline
\vpad
  10C J104320+585621 & 3.27 & 0.11 & 1.08 & 0.35 & 1 &   & 97.0 &   & 479 & $1.86 \times 10^{25}$ & L & P &   & FRII \\
  10C J104328+590312 & 0.92 & 0.09 & 0.48 & 0.24 & 2 & < & 1.5 & < & 6 & $4.32 \times 10^{23}$ & L & P &   & FR0u\\
  10C J104344+591503 & 2.90 & 0.17 & 0.31 & 0.91 & 2 &   & 1.7 &   & 13 & $1.63 \times 10^{25}$ & L & F &   & FR0u\\
  10C J104428+591540 & 0.81 & 0.09 & -0.58 & 0.36 & 1 &   & 2.5 &   & 13 & $5.42 \times 10^{22}$ & H & U &   & FR0g\\
  10C J104441+591949 & 0.61 & 0.09 & -0.08 & 1.30 & 2 &   & 1.6 &   & 13 & $2.04 \times 10^{24}$ & L & U &   & FR0g\\
  10C J104451+591929 & 1.07 & 0.10 & 0.93 & 0.96 & 2 &   & 3.3 &   & 26 & $4.54 \times 10^{25}$ & L & S &   & FR0u\\
  10C J104528+591328 & 1.78 & 0.12 & 0.82 & 2.31 & 1 & < & 0.3 & < & 2 & $4.27 \times 10^{26}$ & H & S &   & FR0c\\
  10C J104539+585730 & 0.93 & 0.09 & 0.36 & 0.39 & 1 &   & 146.0 &   & 772 & $9.45 \times 10^{23}$ & L & F &   & FRII  \\
  10C J104551+590838 & 0.61 & 0.08 & 0.11 & 0.75 & 2 & < & 0.2 & < & 1 & $1.24 \times 10^{24}$ & H & F &   & FR0u\\
  10C J104624+590447 & 1.10 & 0.09 & 1.16 & 1.86 & 2 & < & 2.7 & < & 22 & $5.23 \times 10^{26}$ & L & S &   & FR0u\\
  10C J104630+582748 & 5.34 & 0.23 & 0.99 & 0.12 & 1 &   & 110.0 &   & 233 & $2.10 \times 10^{24}$ & L & S &   & FR0u\\
  10C J104633+585816 & 0.72 & 0.08 & -0.41 & 0.85 & 1 &   & 2.2 &   & 17 & $3.97 \times 10^{23}$ & H & U &   & FR0g\\
  10C J104648+590956 & 0.78 & 0.09 & -0.11 & 5.88 & 4 & < & 0.5 & < & 3 & $2.70 \times 10^{25}$ & H & F &   & FR0u\\
  10C J104700+591903 & 2.30 & 0.13 & 0.95 &   &   & < & 1.5 &   &   &   & H & S &   & FR0c\\
  10C J104710+582821 & 12.23 & 0.63 & -0.46 & 0.59 & 2 & < & 3.9 & < & 26 & $2.98 \times 10^{24}$ & H & R &   & FR0u\\
  10C J104718+585119 & 1.04 & 0.09 & 0.53 & 0.64 & 2 &   & 35.0 &   & 240 & $5.05 \times 10^{24}$ & H & P &   & FRI \\
  10C J104719+582114 & 45.71 & 2.29 & 0.18 & 1.22 & 1 & < & 0.7 & < & 6 & $3.15 \times 10^{26}$ & H & P &   & FR0u\\
  10C J104733+591244 & 1.47 & 0.11 & 0.11 &   &   & < & 1.1 &   &   & $1.19 \times 10^{25}$ & H & F &   & FR0u\\
  10C J104737+592028 & 0.53 & 0.07 & 0.78 & 0.79 & 2 & < & 1.4 & < & 10 & $9.12 \times 10^{24}$ & L & P &   & FR0c\\
  10C J104741+584811 & 0.63 & 0.08 & 1.09 &   &   &   & 20.0 &   &   &   & L & S &   & FR0u\\
  10C J104742+585318 & 0.91 & 0.09 & 0.59 & 0.58 & 2 &   & 32.0 &   & 211 & $4.41 \times 10^{24}$ & L & P &   & FRII \\
  10C J104751+574259 & 1.31 & 0.17 & 0.46 & 1.63 & 2 & < & 2.0 & < & 17 & $4.16 \times 10^{25}$ &   & F &   & FR0u\\
  10C J104802+574117 & 1.01 & 0.17 & 0.90 & 1.38 & 2 & < & 2.0 & < & 17 & $9.26 \times 10^{25}$ & L & S &   & FR0u\\
  10C J104822+582436 & 2.54 & 0.18 & 0.82 & 1.23 & 2 & < & 1.7 & < & 14 & $1.39 \times 10^{26}$ & L & S &   & FR0c\\
  10C J104824+583029 & 2.92 & 0.19 & 0.40 & 0.76 & 4 & < & 4.0 & < & 29 & $1.45 \times 10^{25}$ & L & F &   & FR0u\\
  10C J104826+584838 & 0.57 & 0.08 & 0.04 & 0.96 & 2 & < & 1.3 & < & 10 & $1.53 \times 10^{24}$ & L & S &   & FR0u\\
  10C J104836+591846 & 0.57 & 0.07 & 0.63 & 1.94 & 2 &   & 3.9 &   & 33 & $4.67 \times 10^{25}$ & H & P &   & FR0u\\
  10C J104844+582309 & 2.87 & 0.19 & 0.52 & 0.86 & 2 &   & 22.0 &   & 169 & $2.75 \times 10^{25}$ & L & S &   & FR0u\\
  10C J104849+571417 & 1.82 & 0.13 & 0.71 & 0.61 & 2 & < & 5.4 & < & 36 & $1.40 \times 10^{25}$ & L & S & N & FR0u\\
  10C J104856+575528 & 1.28 & 0.17 & 0.88 & 4.75 & 2 &   & 37.0 &   & 238 & $2.04 \times 10^{27}$ &   & S &   & FRI \\
  10C J104857+584103 & 1.48 & 0.10 & 0.81 & 0.68 & 2 & < & 1.0 & < & 7 & $1.97 \times 10^{25}$ &   & S &   & FR0c\\
  10C J104906+571156 & 1.86 & 0.14 & 0.83 &   &   &   & 37.0 &   &   &   &   & S &   & FRII \\
  10C J104918+582801 & 6.64 & 0.36 & 0.54 & 2.30 & 1 & < & 1.3 & < & 11 & $5.73 \times 10^{26}$ & H & S &   & FR0c\\
  10C J104927+583830 & 0.75 & 0.08 & 0.24 &   &   & < & 30.0 &   &   &   &   & S &   & FR0u\\
  10C J104934+570613 & 2.09 & 0.14 & 0.78 & 1.13 & 2 &   & 25.0 &   & 205 & $8.08 \times 10^{25}$ & L & S & Y & FR0u\\
  10C J104939+583530 & 21.12 & 1.06 & 0.93 & 0.97 & 1 & < & 1.4 & < & 11 & $9.30 \times 10^{26}$ & H & S &   & FR0c\\
  10C J104943+571739 & 1.17 & 0.09 & -0.08 & 0.59 & 1 &   & 1.4 &   & 9 & $8.52 \times 10^{23}$ & H & F & Y & FR0g\\
  10C J104954+570456 & 5.64 & 0.30 & -0.71 & 0.53 & 1 & < & 10.9 & < & 69 & 5.48 $\times 10^{23}$ & H & R & Y & FR0u\\
  10C J105000+585227 & 0.57 & 0.08 & -0.38 & 1.26 & 4 &   & 12.2 &   & 102 & 6.90 $\times 10^{23}$ &   & U &   & FR0g\\
  10C J105007+572020 & 1.18 & 0.08 & 0.28 & 1.22 & 2 & < & 1.0 & < & 8 & 1.12 $\times 10^{25}$ &   & F & Y & FR0u\\
  10C J105007+574251 & 0.72 & 0.07 & 0.24 & 0.88 & 2 &   & 11.4 &   & 88 & 3.01 $\times 10^{24}$ &   & P &   & FR0u\\
  10C J105009+570724 & 1.05 & 0.10 & 0.67 &   &   &   & 25.0 &   &   &   &   & S & Y & FRI \\
  10C J105020+574048 & 0.82 & 0.08 & 0.71 & 0.71 & 2 &   & 8.0 &   & 58 & 8.75 $\times 10^{24}$ & L & S &   & FR0u\\
  10C J105028+574522 & 0.53 & 0.07 & -0.28 & 0.07 & 1 & < & 12.3 & < & 17 & 3.16 $\times 10^{21}$ & L & U &   & FR0u\\
  10C J105034+572922 & 1.20 & 0.09 & -0.52 & 1.12 & 3 & < & 0.8 & < & 7 & 7.48 $\times 10^{23}$ &   & U & N & FR0u\\
  10C J105039+572339 & 1.08 & 0.08 & 0.76 & 1.44 & 1 & < & 0.4 & < & 3 & 7.06 $\times 10^{25}$ & H & P & Y & FR0c\\
  10C J105039+574200 & 0.86 & 0.08 & 0.14 & 0.86 & 2 &   & 11.3 &   & 87 & 2.55 $\times 10^{24}$ &   & S &   & FR0u\\
  10C J105039+585118 & 1.00 & 0.10 & 0.75 & 0.37 & 1 &   & 47.0 &   & 243 & 2.72 $\times 10^{24}$ & L & S &   & FRII \\
  10C J105040+573308 & 0.89 & 0.09 & 0.00 &   &   & < & 0.8 &   &   &   &   & S & Y & FR0u\\
  10C J105042+575233 & 3.51 & 0.22 & 1.00 & 1.24 & 2 &   & 17.0 &   & 142 & 3.46 $\times 10^{26}$ &   & S &   & FR0u\\
  10C J105050+580200 & 5.67 & 0.32 & -0.18 & 0.68 & 2 & < & 3.0 & < & 21 & 4.10 $\times 10^{24}$ & H & F &   & FR0u\\
  10C J105053+583233 & 22.94 & 1.16 & 0.79 & 4.53 & 4 & < & 3.6 & < & 24 & 2.22 $\times 10^{28}$ & H & S &   & FR0u\\
  10C J105054+580943 & 1.45 & 0.19 & 0.21 &   &   & < & 2.2 &   &   &   &   & P &   & FR0u\\
  10C J105058+573356 & 0.96 & 0.09 & -0.65 &   &   & < &   &   &   &   & L & U &   & FR0g\\
  10C J105104+574456 & 1.17 & 0.09 & -0.71 &   &   &   &   &   &   &   &   & U &   & FR0g\\
  10C J105104+575415 & 8.14 & 0.43 & 0.34 & 1.67 & 1 & < & 1.6 & < & 14 & 1.78 $\times 10^{26}$ & H & S &   & FR0u\\
  10C J105107+575752 & 2.89 & 0.19 & 0.64 & 0.69 & 2 &   & 17.0 &   & 121 & 2.32 $\times 10^{25}$ & L & S &   & FR0u\\
  10C J105115+573552 & 0.53 & 0.08 & 0.65 &   &   & < & 1.3 &   &   &   & H & S & Y & FR0c\\
  10C J105121+582648 & 3.43 & 0.22 & 1.06 & 1.44 & 4 &   & 20.0 &   & 169 & 6.14 $\times 10^{26}$ &   & S &   & FRI  \\
  10C J105122+570854 & 1.26 & 0.10 & 0.87 & 1.13 & 3 & < & 0.6 & < & 5 & 6.64 $\times 10^{25}$ &   & S & Y & FR0c\\
  10C J105122+584136 & 1.45 & 0.16 & 0.49 & 1.68 & 2 & < & 1.9 & < & 16 & 5.33 $\times 10^{25}$ &   & S &   & FR0u\\
  10C J105122+584409 & 1.38 & 0.15 & 0.34 & 1.79 & 2 &   & 11.1 &   & 94 & 3.48 $\times 10^{25}$ & H & P &   & FR0u\\
  10C J105128+570901 & 2.25 & 0.14 & 0.70 & 0.54 & 1 &   & 30.0 &   & 191 & 1.25 $\times 10^{25}$ & L & S & Y & FRII  \\
  
  \hline\end{tabular}
  \end{table*}

\begin{table*}
\contcaption{}
\small\rm
\medskip
\centering
\renewcommand{\tabcolsep}{1.6mm}
\begin{tabular}{lddddccdcdcrrccc@{}}\hline
\vpad
ID & \dhead{$S_{15.7~\rm GHz}$} & \dhead{$\sigma\_S_{15.7~\rm GHz}$} & \dhead{$\alpha^{15.7}_{1.4}$} & z & $z$            & \multicolumn{2}{c}{ang. size} & \multicolumn{2}{c}{lin. size} & $L_{1.4~\rm GHz}$ & HERG & spec & VLBI & morph.  \\
   & \dhead{(mJy)}   & \dhead{(mJy)}              &                               &   & flag & \multicolumn{2}{c}{(arcsec)}    & \multicolumn{2}{c}{(kpc)}       & W / Hz          & class    & type &      & class  \\
 (1) & (2) & (3) & (4) & (5) & \multicolumn{2}{c}{(6)} & \multicolumn{2}{c}{(7)} & (8) & (9) & (10) & (11) & (12) & (13) \\     
\hline
\vpad
  10C J105132+571114 & 2.92 & 0.16 & 0.60 & 0.32 & 1 &   & 52.0 &   & 241 & 3.69 $\times 10^{24}$ & L & P & Y & FRII  \\
  10C J105136+572944 & 1.12 & 0.08 & 0.38 & 4.91 & 4 &   & 2.9 &   & 18 & 2.32 $\times 10^{26}$ & H & F & Y & FR0  \\
  10C J105138+574957 & 0.67 & 0.08 & 0.55 &   &   &   &   &   &   &   &   & S &   & FR0u\\
  10C J105139+580757 & 2.49 & 0.20 & 1.08 &   &   &   & 36.0 &   &   &   &   & S &   & FRI \\
  10C J105142+573447 & 0.79 & 0.08 & 0.02 & 0.73 & 2 & < & 0.3 & < & 2 & 1.15 $\times 10^{24}$ & H & F & Y & FR0* \\
  10C J105142+573557 & 1.95 & 0.12 & 0.37 & 1.44 & 2 & < & 0.4 & < & 3 & 3.55 $\times 10^{25}$ & H & S & Y & FRII* \\
  10C J105144+573313 & 0.98 & 0.09 & 1.06 &   &   &   & 124.3 &   &   &   & L & S &   & FR0u\\
  10C J105148+573245 & 0.52 & 0.07 & 0.24 & 0.99 & 1 &   & 114.0 &   & 911 & 2.84 $\times 10^{24}$ & H & F & Y & FR0u\\
  10C J105206+574111 & 3.35 & 0.18 & 0.50 & 0.46 & 1 & < & 0.9 & < & 5 & 7.44 $\times 10^{24}$ & H & S & Y & FR0c\\
  10C J105215+581627 & 1.13 & 0.14 & 0.43 & 0.91 & 2 & < & 2.3 & < & 18 & 9.26 $\times 10^{24}$ &   & S &   & FR0u\\
  10C J105220+585051 & 2.03 & 0.16 & 0.80 & 1.94 & 2 & < & 2.3 & < & 19 & 3.02 $\times 10^{26}$ &   & S &   & FR0u\\
  10C J105225+573323 & 0.58 & 0.08 & 0.90 & 0.61 & 1 &   & 18.0 &   & 121 & 7.66 $\times 10^{24}$ & L & S & N & FR0u\\
  10C J105225+575507 & 22.45 & 1.13 & 0.11 & 4.87 & 4 & < & 1.2 & < & 8 & 1.50 $\times 10^{27}$ &   & P &   & FR0u\\
  10C J105237+573058 & 5.16 & 0.19 & 1.05 &   &   &   & 52.0 &   &   &   & L & S & Y & FRII \\
  10C J105240+572322 & 0.54 & 0.06 & 0.50 & 1.11 & 3 & < & 1.3 & < & 11 & 8.34 $\times 10^{24}$ & H & F & Y & FR0u\\
  10C J105243+574817 & 1.00 & 0.08 & 0.26 & 0.67 & 2 & < & 0.9 & < & 6 & 2.46 $\times 10^{24}$ & L & P & Y & FR0u\\
  10C J105327+574546 & 0.81 & 0.08 & 0.89 & 0.82 & 3 &   & 30.0 &   & 227 & 2.12 $\times 10^{25}$ &   & S & N & FR0u\\
  10C J105341+571951 & 0.55 & 0.07 & 0.42 & 0.91 & 3 & < & 0.6 & < & 5 & 4.39 $\times 10^{24}$ & L & S & Y & FR0u\\
  10C J105342+574438 & 1.80 & 0.11 & 0.59 & 0.83 & 2 &   & 32.0 &   & 243 & 1.92 $\times 10^{25}$ & L & S & Y & FR0/I  \\
  10C J105400+573324 & 0.67 & 0.07 & 0.62 & 1.49 & 3 & < & 1.0 & < & 8 & 2.96 $\times 10^{25}$ & H & S & Y & FR0c\\
  10C J105425+573700 & 25.47 & 1.01 & 0.83 & 0.32 & 1 &   & 74.0 &   & 346 & 6.20 $\times 10^{25}$ & H & S & Y & FRII \\
  10C J105437+565922 & 2.20 & 0.15 & 0.98 &   &   &   & 144.0 &   &   &   &   & P &   & FR0u\\
  10C J105441+571640 & 0.78 & 0.07 & 0.63 &   &   &   & 22.0 &   &   &   &   & S & Y & FRII  \\
  10C J105510+574503 & 0.73 & 0.08 & 0.15 & 1.14 & 2 & < & 11.5 & < & 95 & 3.96 $\times 10^{24}$ & L & F &   & FR0u\\
  10C J105515+573256 & 0.63 & 0.08 & 0.94 &   &   &   & 30.0 &   &   &   &   & S & N & FRI/0  \\
  10C J105520+572237 & 0.56 & 0.07 & 0.23 & 0.16 & 4 & < & 1.4 & < & 4 & 6.00 $\times 10^{22}$ & H & F &   & FR0u\\
  10C J105527+571607 & 0.83 & 0.08 & 0.56 & 0.49 & 1 & < & 4.6 & < & 27 & 2.42 $\times 10^{24}$ & L & S &   & FR0u\\
  10C J105535+574636 & 2.68 & 0.16 & -0.03 & 2.89 & 2 & < & 1.8 & < & 14 & 4.34 $\times 10^{25}$ & H & F &   & FR0u\\
  10C J105550+570407 & 1.19 & 0.14 & -0.33 & 0.49 & 1 & < & 4.7 & < & 29 & 2.83 $\times 10^{23}$ & L & U &   & FR0u\\
  10C J105604+570934 & 5.47 & 0.31 & 0.93 & 4.88 & 4 & < & 1.3 & < & 8 & 1.12 $\times 10^{28}$ &   & S &   & FR0c\\
  10C J105627+574221 & 1.09 & 0.16 & -0.05 & 1.64 & 2 & < & 3.4 & < & 29 & 6.15 $\times 10^{24}$ & H & U &   & FR0u\\
  10C J105653+580342 & 1.95 & 0.19 & -0.42 & 0.60 & 1 &   & 11.3 &   & 76 & 5.39 $\times 10^{23}$ & L & F &   & FR0g\\
  10C J105716+572314 & 6.42 & 0.37 & 1.25 &   &   &   & 18.0 &   &   &   &   & S &   & FR0/I  \\
\hline\end{tabular}
\vpad
\begin{flushleft}
 Notes:\\
(1) source name; 
(2) 15.7-GHz flux density (mJy), see Section 2 in Paper I for details;
(3) uncertainty in the 15.7-GHz flux density (mJy);
(4) radio spectral index, see Section 4.2 in Paper I for details, median uncertainty = $\pm 0.01$;
(5) redshift, see Section 4.3 in Paper II for details;
(6) redshift flag indicating the origin of the redshift value in (5), 1 = spectroscopic redshift (references for individual sources are listed in Table A2 in Paper II), 2 -- 4 = photometric redshift; 2 = from \citet{2013MNRAS.428.1958R}, mean value of $(z_{\rm phot} - z_{\rm spec})/(1 + z_{\rm phot}) = 3.5$~per cent, 3 =  from \citet{2012ApJS..198....1F}, median error in fit = 12 per cent, 4 = from fitting described in Paper II, median error in fit = 1 per cent;
(7) angular size (arcsec), see Section 6.4 in Paper II. Upper limits are marked with `$<$', errors in measured values are 10 per cent;
(8) linear size (kpc), see Section 6.4 in Paper II. Upper limits are marked with `$<$';
(9) 1.4-GHz luminosity (W / Hz), see Section 6.3 in Paper II for details;
(10) excitation classification. H = probable HERG, L = probable LERG, blank = not enough information to classify. See Section~\ref{section:HERG_overall_class};
(11) radio spectral classification, S = steep, R = rising, P = peak, U = upturn, F = flat. The criteria for the classifications are given in Table~\ref{tab:spec_class};
(12) VLBI classification, Y = detected by the VLBI observations, N = not detected by the VLBI observations, blank = source is outside the area covered by the VLBI observations. See Section~\ref{section:VLBI} and \citet{2014MNRAS.440...40W} for details;
(13) morphological classification. The criteria are listed in Table~\ref{tab:morph_class}.
\end{flushleft}
\end{table*}

\subsection{Very Long Baseline Interferometry}\label{section:VLBI}

For 30 of the sources studied in this paper it is possible to get more information about their structure from Very Long Baseline Interferometry (VLBI) observations made over part of the field by \citet{2013A&A...551A..97M} using the Very Long Baseline Array (VLBA). The 10C sources in the region covered by these observations are discussed in detail by \citet{2014MNRAS.440...40W} who show that 24 out of the 30 10C sources in the field have a compact core (see Table~\ref{tab:properties}), as expected for radio galaxies. All 12 of the HERGs in this sample have a compact core, while two out of the nine LERGs do not, providing further support for the idea that the 10C HERGs tend to be more core-dominated than the LERGs. Only three of the 24 sources are found to be compact on scales $<$10~mas, suggesting that the majority of the sources have some extended emission on scales between 10~mas and 6~arcsec. 

By contrast, in their study of the shallower 15-GHz 9C survey (limiting flux density 25 mJy), \citet{2006MNRAS.367..323B} found that none of the 16 sources with spectra peaking above 2.5~GHz in the sample they selected were resolved, even with a resolution of 3~mas. They concluded from this result and a variability study of 9C sources \citep{2006MNRAS.370.1556B} that a large proportion of the `GPS' sources in their sample are actually beamed sources. As all but three of the 10C sources with VLBI information are extended on scales $>10$~mas, this is unlikely to be the case in our sample.

\subsection{Properties of the different morphological classifications}

\begin{figure}
\centerline{\includegraphics[width=\columnwidth]{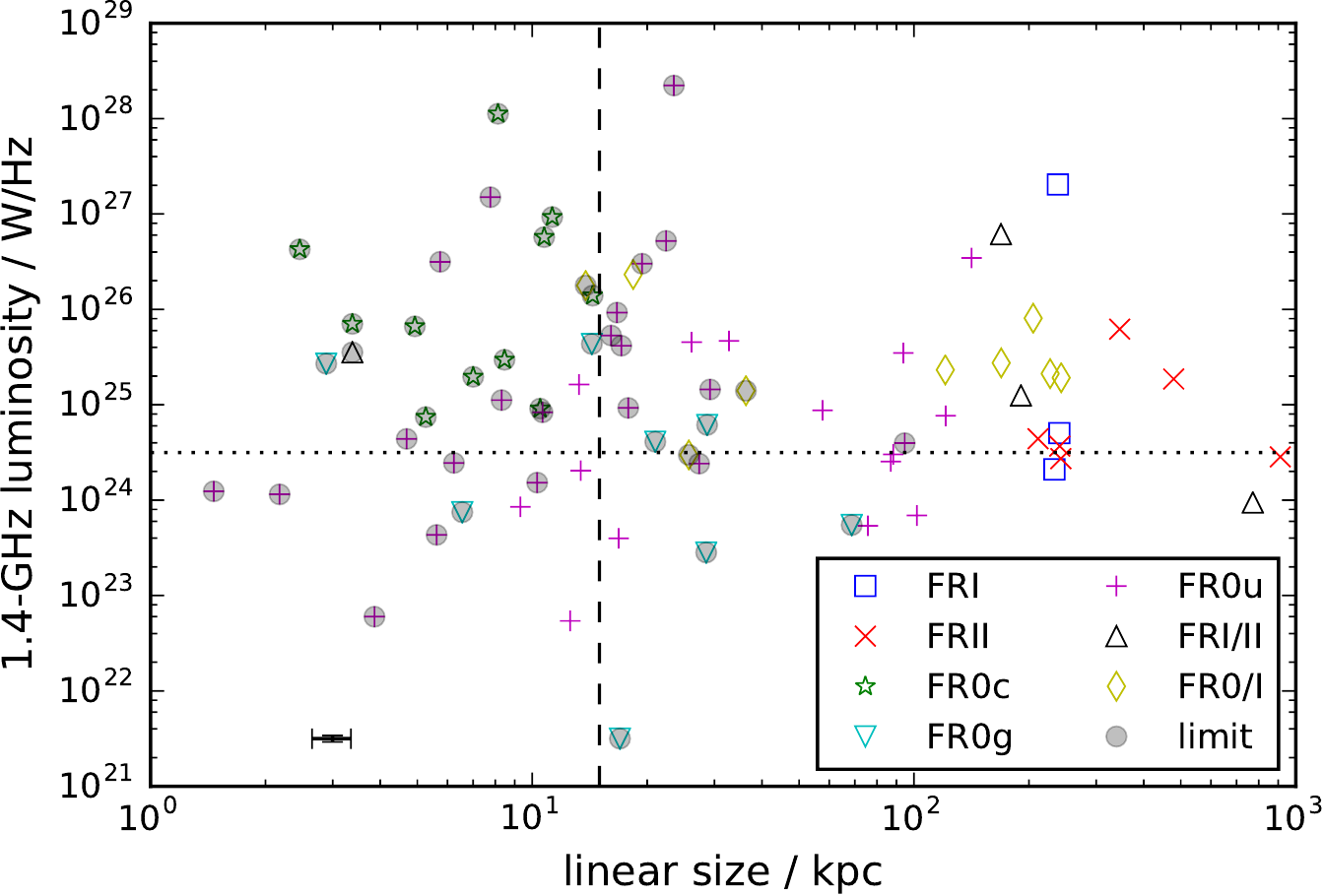}}
\caption{ 1.4-GHz luminosity as a function of linear size, with sources split according to their morphological classification. Sources with upper limits on their linear size are marked with grey circles, and could move to the left. The vertical dashed line is at $d = 15$~kpc, CSS sources are typically smaller than this value. The horizontal dotted line is at ${\rm log(}L_{1.4~\rm GHz}  / \rm W \, Hz^{-1}) = 24.5$, the typical luminosity divide between FRI and FRII sources. For clarity the individual error bars have been omitted but a point representing the median uncertainties in luminosity and linear size is included in the bottom left corner. The 77 sources with redshift information available are included in this plot. }\label{fig:lum14_linsize}
\end{figure}

The properties of all sources in this sample are summarised in Table~\ref{tab:properties}. The 1.4-GHz luminosities of the 10C sources are shown as a function of linear size in Fig.~\ref{fig:lum14_linsize} for the 77 sources with redshift values available. In the left panel the sources are split according to their morphological class, and in the right panel sources with an upper limit on linear size are shown separately. 
Note that six of the FRI and FRII sources are not included in this diagram; they do not have redshift values available because they were classified as `confused' when matching to the multi-wavelength catalogue (see Section 3.2 in Paper II). FRI and FRII sources are found on both sides of the typical dividing luminosity $L_{1.4~\rm GHz} = 10^{24.5}~\rm W \,Hz^{-1}$. The FR0c (CSS) sources are generally luminous, with all 11 with redshifts available having $L_{1.4~\rm GHz} > 10^{24.5}~\rm W \, Hz^{-1}$ and a median luminosity of $L_{1.4~\rm GHz} = 7 \times 10^{25}~\rm W \, Hz^{-1}$. The FR0g (GPS) sources tend to be less luminous, with a median luminosity of $L_{1.4~\rm GHz} = 3 \times 10^{24}~\rm W \, Hz^{-1}$. This result is partly due to a selection effect as by definition all the FR0c sources have steep spectra and all the FR0g sources have flat or rising spectra; consequently as the sources are selected at 15.7~GHz but the luminosities are calculated at 1.4~GHz, the FR0c sources will tend to have higher luminosities than the FR0g sources. However, this effect is still visible when comparing 15.7-GHz luminosities, with the FR0c sources having a median luminosity of $L_{15.7~\rm GHz} = 1.1 \times 10^{25}~\rm W \, Hz^{-1}$, while the FR0g sources have a median of $L_{15.7~\rm GHz} = 6.3 \times 10^{24}~\rm W \, Hz^{-1}$.

\begin{figure}
\centerline{\includegraphics[width=\columnwidth]{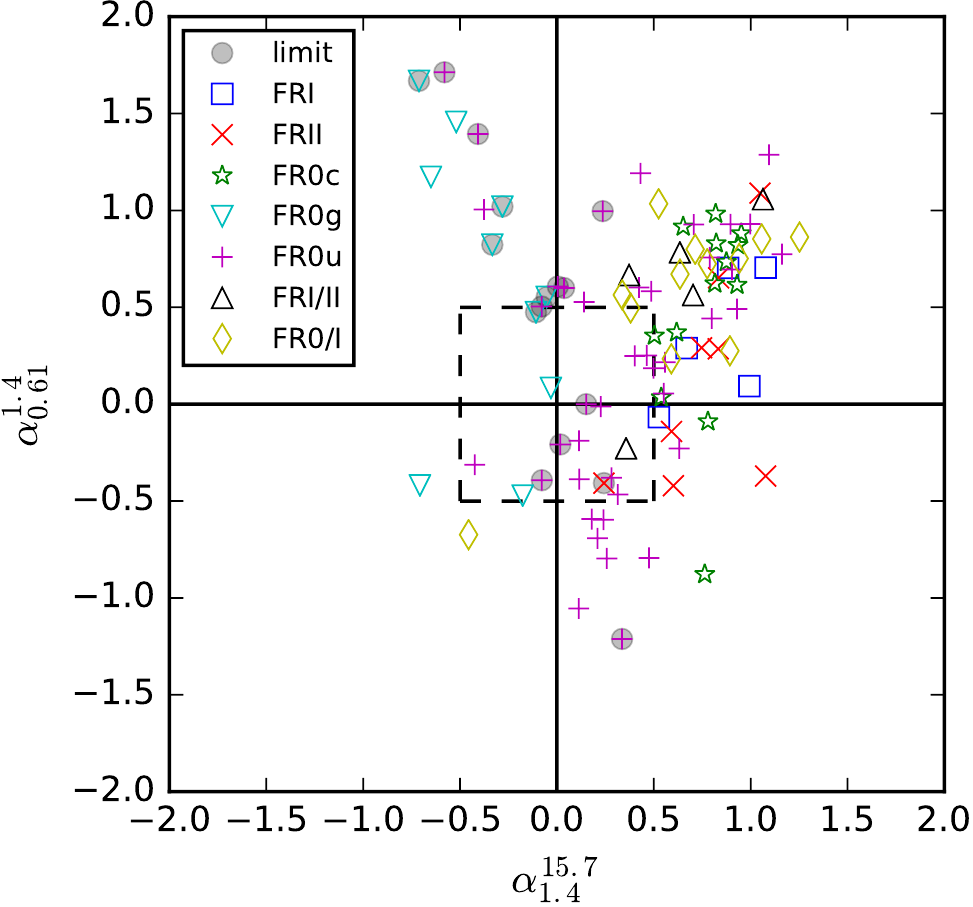}}
\caption{Radio colour--colour diagram with sources split according to their morphological classification. The dashed square encloses the region with $-0.5 < \alpha^{15.7}_{1.4} < 0.5$ and $-0.5 < \alpha^{1.4}_{0.61} < 0.5$ where flat-spectrum sources lie. The solid lines are at $\alpha=0$ and indicate the divisions between the different spectral classes listed in Table~\ref{tab:spec_class}. Sources with an upper limit on $\alpha^{1.4}_{0.61}$ are marked by grey circles, these sources could move down on this plot (note that all the sources are detected at both 15.7 and 1.4~GHz). The median uncertainties in the spectral indices are $\pm 0.01$. All 96 sources in the sample are included in this figure.  }\label{fig:colour-colour-morph}
\end{figure}

The different morphological classes are shown on a radio colour--colour diagram in Fig.~\ref{fig:colour-colour-morph}. We expect the FRI, FRII and FRI/II sources to have steep spectra as they display powerful extended emission, and while this is the case for nine out of the 18 sources, six have flat or peaked spectra. The three sources with peaked spectra, which are flat at lower frequencies but steep at higher frequencies, are particularly surprising, as we would expect the reverse trend, with spectra flattening at higher frequencies as flat-spectrum emission from the cores contribute to the total emission. All but one of the FR0/I sources (which are extended but do not display evidence for two components) have steep spectra, showing that these sources are dominated by their extended emission. By definition, the FR0g (GPS) sources all have flat or rising spectra between 1.4 and 15.7 GHz. Several of these sources appear to have steep lower-frequency spectra, however for most of the sources this is because the 610 MHz data point used for these sources is an upper limit (indicated by a grey circle), so the spectral index between 610 MHz and 1.4 GHz is almost certainly significantly flatter than the values shown in the figure. For three sources, however, this is not the case, meaning that these sources have an unusual `inverted' spectral shape. This could indicate that the cores of these sources are dominating at higher frequencies, or be because these sources are variable as the observations at the different frequencies were not simultaneous. (One of these sources is discussed in more detail in \citealt{2014MNRAS.440...40W}). Although all of the FR0c (CSS) sources have steep spectra between 1.4 and 15.7 GHz, two of the 13 sources have flat spectra between 610 MHz and 1.4 GHz, showing that the spectra turn over at $\lesssim 1$~GHz. The unclassified FR0 sources display a range of different spectral shapes, indicating that, unsurprisingly, this is a mixed population of different source types. Sources in the lower left quadrant are flat or rising from 610~MHz through to 15.7~GHz, indicating that they peak above 15~GHz. These sources could therefore be examples of `high-frequency peakers', which have spectra which peak at a few tens of GHz (see e.g.\ \citealt{2000A&A...363..887D,2010MNRAS.408.1187H}).

\subsection{Comparison between morphological classes and HERG/LERG classification}

\begin{table}
\caption{Comparison of morphological and excitation classifications of 10C sources.}\label{tab:morph_spec}
\medskip
\centering
\begin{tabular}{l|rrr|r}\hline\vpad
       & HERG & LERG & Unclass. & Total \\\hline
FRI    & 1    & 1    & 3        & 5 \\
FRII   & 2    & 5    & 1        & 8\\
FRI/II & 1    & 2    & 2        & 5\\
FR0/I  & 3    & 6    & 4        & 13\\
FR0c   & 8    & 2    & 3        & 13\\
FR0g   & 6    & 2    & 2        & 10 \\
FR0u   & 11   & 17   & 14       & 42 \\\hline
Total  & 32   & 35   & 29       & 96 \\\hline
\end{tabular}
\end{table}

The morphological and excitation classifications are compared in Table~\ref{tab:morph_spec}. There is some indication that more FR0c and FR0g sources may be HERGs than LERGs. CSS and GPS sources are the most compact source types studied here, so this is consistent with the conclusion from Section~\ref{section:HERG_overall} that the HERGs in the 10C sample are smaller and in some cases more core-dominated than the LERGs. However, due to the small numbers of sources involved and the large proportion of sources which do not have excitation classifications or are in the FR0u category, this result may not be significant. The tendency for the LERGs in this sample to be more extended than the HERGs supports the scenario in which HERGs emit efficiently across the whole electromagnetic spectrum while LERGs emit most of their energy in kinetic form, causing them to appear more extended in the radio (e.g.\ \citealt{2012MNRAS.421.1569B}).

%------------------------------------------------------------------------------%
\section{Comparison with the population at higher flux densities}\label{section:other-work}

The AT20G survey \citep{2010MNRAS.402.2403M} was carried out with the Australia Telescope Compact Array at 20~GHz and surveyed the whole southern sky, detecting nearly 6000 radio sources with $S_{20~\rm GHz} > 40$~mJy. As the AT20G survey probes significantly higher flux densities than the 10C survey, comparing the two samples allows us to look at how high-frequency selected radio sources change over a wide range of flux densities. We have already shown in Paper I how the spectral indices of this population change significantly with 15.7-GHz flux density: from a population dominated by flat spectrum sources at high flux densities (> 100 mJy), to a population where the majority of the sources have steep spectra between $1 \lesssim S_{15~\rm GHz}/{\rm mJy} \lesssim 100$, before returning to a flat-spectrum dominated population below 1 mJy.

\begin{figure}
\centerline{\includegraphics[width=\columnwidth]{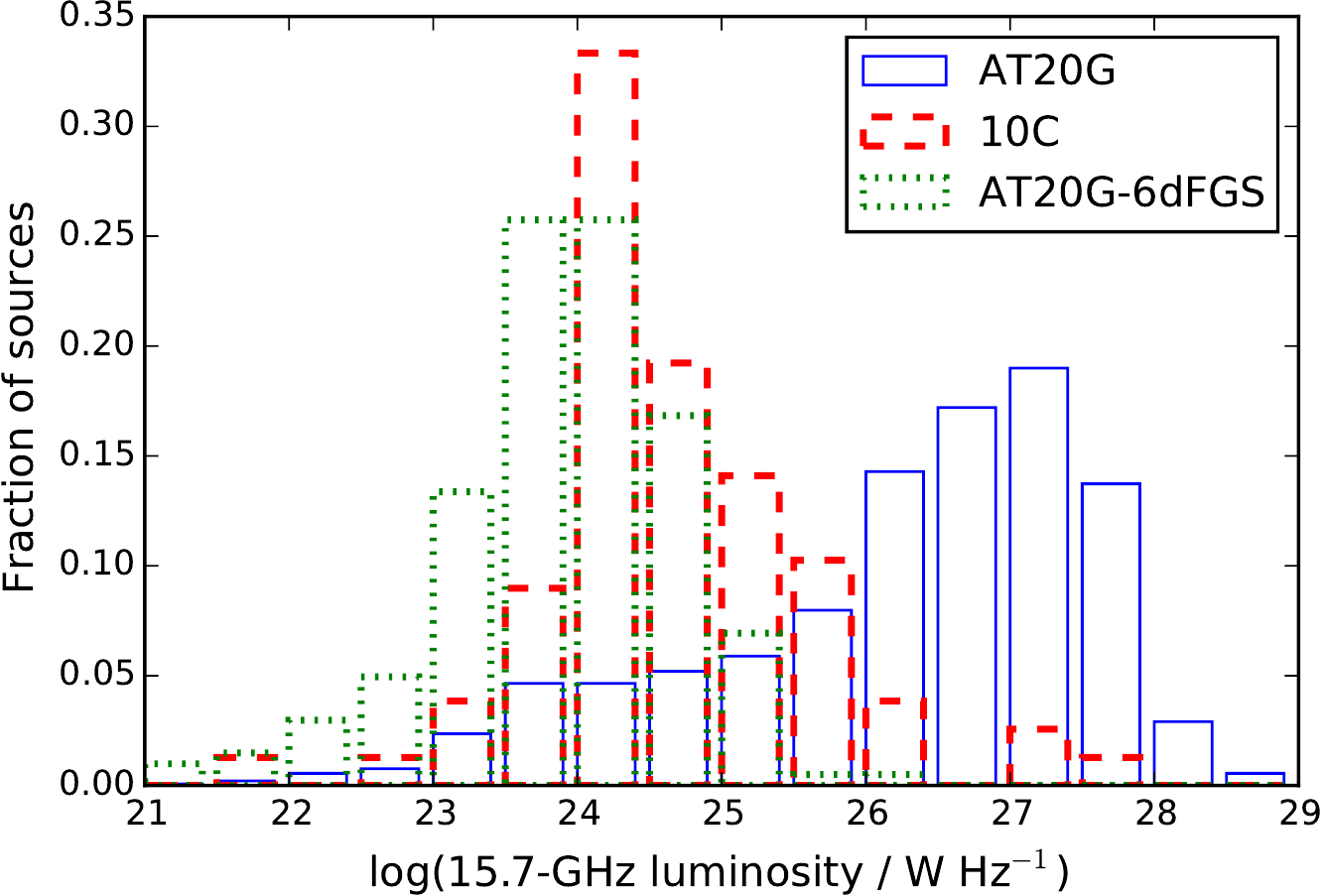}}
\caption{Luminosity distribution of the AT20G \citep{2011MNRAS.417.2651M}, AT20G-6dFGS \citep{2014MNRAS.438..796S} and 10C samples. The histograms are normalised by the total number of sources with redshifts available in each sample. For the AT20G and AT20G-6dFGS samples the luminosities are corrected from 20 to 15.7~GHz using the spectral index of each source. The 10C sample plotted here contains the 77 sources in the sample discussed in this paper which redshifts available.}\label{fig:AT20G-lum}
\end{figure}

\begin{figure}
\centerline{\includegraphics[width=\columnwidth]{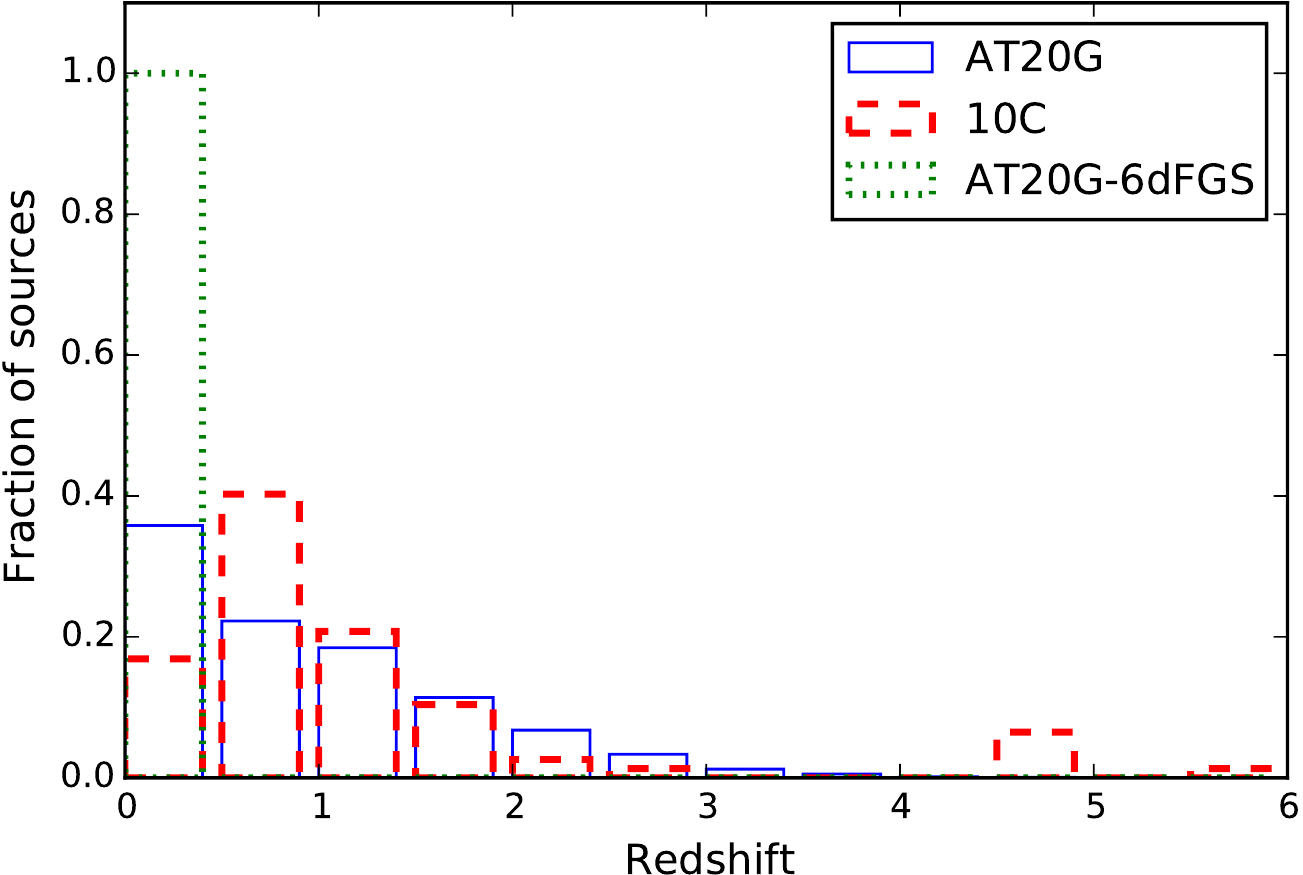}}
\caption{Redshift distribution of the AT20G \citep{2011MNRAS.417.2651M}, AT20G-6dFGS \citep{2014MNRAS.438..796S} and 10C samples. The histograms are normalised by the total number of sources with redshifts available in each sample. The 10C sample plotted here contains the 77 sources in the sample discussed in this paper which have redshifts available. There are 1464 sources in the AT20G sample and 202 and the AT20G-6dFGS sample.}\label{fig:AT20G-z}
\end{figure}

The optical properties of the 4932 AT20G sources outside the galactic plane were studied by \citet{2011MNRAS.417.2651M} using data from the SuperCOSMOS database. More recently, a sub-sample of 202 nearby (median redshift of 0.058) AT20G sources with counterparts in the 6dF Galaxy Survey (6dFGS) were investigated by \citet{2014MNRAS.438..796S} (we will refer to this sub-sample as the AT20G-6dFGS sample). Here, we compare these two samples (the full AT20G sample and the low-redshift AT20G-6dFGS sample) to the 10C sample studied in this paper. The 15.7-GHz luminosity distributions are given in Fig.~\ref{fig:AT20G-lum}, the AT20G luminosities are shifted from 20 to 15.7 GHz using the spectral index of each source ($\alpha^{20}_{0.843}$ is used for those sources with a SUMSS counterpart, and $\alpha^{20}_{1.4}$ is used for sources with an NVSS counterpart). The 10C and AT20G-6dFGS luminosity distributions are quite similar though the 10C distribution is shifted to slightly higher luminosities than the AT20G-6dFGS sample, with a median 15.7-GHz luminosity of $3.9 \times 10^{24}~\rm W \, Hz^{-1}$ compared to $1.0 \times 10^{24}~\rm W \, Hz^{-1}$. The 15.7-GHz luminosity distribution for the full AT20G sample, however, has a different shape and peaks at a significantly higher luminosity ($L_{15.7~\rm GHz} \approx 10^{27} ~\rm W \, Hz^{-1}$) than the other two distributions.  Fig.~\ref{fig:AT20G-z} shows the redshift distributions of the three samples; here the 10C redshift distribution is more similar to that of the full AT20G sample, while the AT20G-6dFGS sources all have low redshifts ($z<0.3$). There is, however, some indication that there are a smaller proportion of low-redshift sources in the 10C sample than in the AT20G sample.  These distributions show that the 10C sources are, in terms of their luminosities, higher-redshift versions of the objects found in the AT20G-6dFGS bright galaxy sample rather than those found in the full AT20G sample.

\begin{figure}
\centerline{\includegraphics[width=\columnwidth]{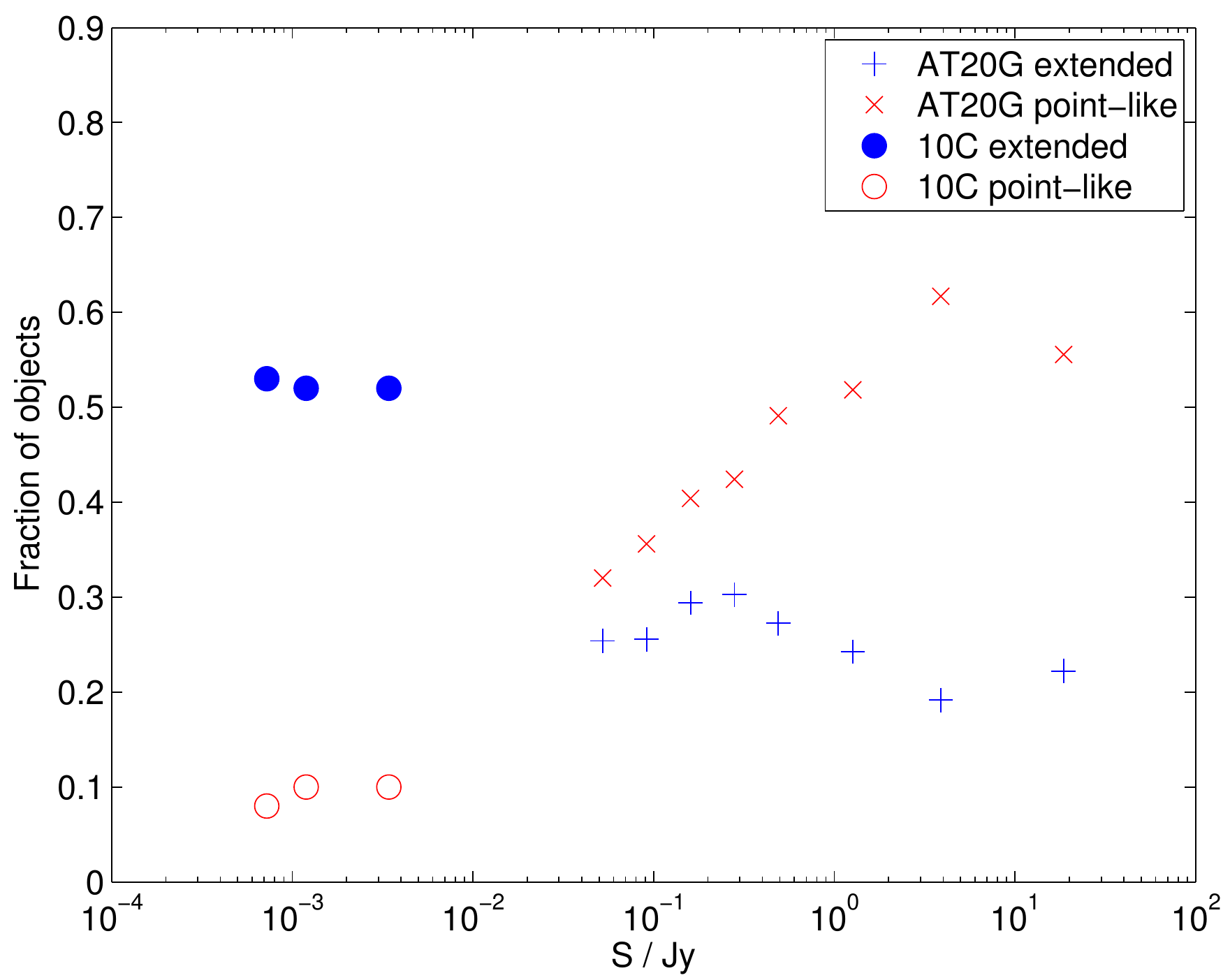}}
\caption{Fraction of objects associated with radio sources classified as extended and point-like in the optical as a function of flux density in the AT20G and 10C surveys (20-GHz flux density for AT20G sources and 15.7-GHz flux density for 10C sources). The 10C sample plotted here consists of the 59 sources with optical compactness information available.}\label{fig:mahony}
\end{figure}

The differences between 10C and the full AT20G sample are also highlighted in the nature of the associated optical objects. \citeauthor{2011MNRAS.417.2651M} investigated the optical compactness of the host galaxies of the AT20G sample and found that the population is dominated by `stellar' sources (which are point-like in the optical) and that the proportion of these sources decreases with decreasing 20-GHz flux density as shown in Fig.~\ref{fig:mahony}.  We find that this trend continues as flux density decreases further -- only nine per cent of the sources with $0.5 < S_{15.7~\rm{GHz}}/~\rm{mJy} < 45$ are point-like in the optical.  As Fig.~\ref{fig:mahony} shows, the population changes steadily from being dominated by point-like objects (i.e.\ quasars) at high flux densities to being dominated by extended objects (i.e. galaxies) at lower flux densities.  In this respect the 10C sample closely resembles the AT20G-6dFGS sample - though the latter by definition does not contain any quasars.

Consistent with the high proportion of optically-compact objects the majority (60 per cent of those with spectra available) of the sources in the AT20G sample are HERGs, while 25 per cent are LERGs.  By contrast 77 per cent of the nearby galaxies in the AT20G-6dFGS sample are LERGs and 23 per cent are HERGs.  The statistics for the 10C sample are less well-defined as about a third are probable HERGs, a third probable LERGs and a third are unclassified; nevertheless it is likely that, despite the similarity in luminosity with the nearby AT20G-6dFGS sample, there is a higher proportion of HERGs in this higher redshift sample. This would be broadly consistent with the \citet{2012MNRAS.421.1569B} picture in which the HERGs show strong cosmic evolution with redshift where as the LERGs show only weak or no evolution.

Comparison of the proportions of sources in the different radio morphological classes indicates that in this respect the 10C sample also resembles the AT20G-6dFGS sample.  A similar proportion of the sources in both samples are classified as FR0 (70 -- 80 per cent of the 10C sample and 70 -- 75 per cent in the AT20G-6dFGS sample); due to the uncertainties in the classification of the FRI and FRII sources in the 10C sample and the relatively small numbers of sources involved it is not possible to make meaningful comparisons of the FRI and FRII populations of the two samples. It is not possible to make comparisons with the radio structures of the sources in the full AT20G sample as they have not been imaged on comparable angular scales.

In summary, the sources in the 10C sample are found at similar redshifts but have lower luminosities than those in the AT20G sample. The powerful sources which dominate the AT20G sample are largely missing from the 10C sample -- instead, the 10C sources seem to be higher-redshift versions of the lower-luminosity, compact radio galaxies found in the AT20G-6dFGS sub-sample.

%------------------------------------------------------------------------------%
\section{Conclusions}\label{section:chap5-conclusions}

We have studied the properties of a sample of 96 radio galaxies selected from the 10C sample in the Lockman Hole. To distinguish between high and low-excitation radio galaxies (HERGs and LERGs) three different methods are used; optical compactness, X-ray observations and mid-infrared colour--colour diagrams. These methods are combined to produce overall HERG and LERG classifications; a total of 32 sources are classified as `probable HERGs', 35 as `probable LERGs' and 29 remain unclassified. 17 sources are also classified using their optical spectra; the spectroscopic classifications agree with the classifications derived here in 94 per cent of cases, showing that this is a reliable way of distinguishing between HERGs and LERGs.

The properties of these HERGs and LERGS are then compared. We find that the HERGs in our sample tend to be found at higher redshifts, have flatter spectra, higher 15.7-GHz flux densities and smaller linear sizes than LERGs. Note however that the relatively large proportion of unclassified sources have the potential to change this result. This result is in contrast to that found for the higher-flux-density AT20G sample, where the HERGs have steeper spectra and are more extended than the LERGs. This suggests that the HERGs in the 10C sample have different properties to their higher-flux-density counterparts, lacking the powerful extended emission typical of FRI and FRII sources, and instead being dominated by their cores.

Low-frequency (610~MHz) radio images, along with radio spectral indices and linear sizes, are used to split the sources into different radio morphological classes. Although 18 sources are found to be FRI or FRII sources and 13 others are significantly extended, the majority of the sample (65 sources) do not display any extended emission on arcsecond scales and are therefore classified as FR0 sources. These FR0 sources are further subdivided into candidate GPS sources (13 sources) and candidate CSS sources (10 sources), while the remaining 42 sources could not be classified. The sub-sample of fainter 10C sources with 15.7-GHz flux density $<$~1~mJy contains a higher proportion of compact FR0 sources (81 per cent), consistent with the result from Paper I that the majority of the faint 10C sources have flat radio spectra.

There is some indication that the CSS and GPS sources are more likely to be HERGs than LERGs, supporting the idea that the 10C HERGs tend to be smaller and in some cases more core-dominated than the LERGs, but a larger proportion of the source must be classified optically into HERGs and LERGs before this conclusion can be confirmed.

By comparing these results to those from both the full AT20G survey \citep{2011MNRAS.417.2651M} and the sub-sample of nearby ($z \sim 0.05$) sources in the AT20G-6dFGS sample \citep{2014MNRAS.438..796S} we have studied the high-frequency extragalactic source population over a wide range in flux density. We find that the fainter 10C sources are not simply higher-redshift versions of the brighter AT20G sources; the two samples are found at similar redshifts, but the 10C sources have significantly lower luminosities. The nature of the optical counterparts to the radio galaxies changes with flux density; at high flux densities most radio sources are associated with quasars, while at low flux densities the optical counterparts are primarily galaxies. We find evidence that the 10C sources may be higher-redshift versions of the lower-luminosity, compact radio galaxies found in the AT20G-6dFGS sub-sample.

This work shows that the faint, flat spectrum radio galaxies which dominate the high-frequency radio sky below 1~mJy are a mixed population of HERGs and LERGs, most of which lack the extended emission typical of FRI or II radio galaxies and have lower luminosities than the higher-flux density sources found in shallower surveys.

%------------------------------------------------------------------------------%
\section*{Acknowledgements}

We thank the referee for their insightful comments which helped to improve this paper. IHW acknowledges a Science and Technology Facilities Council studentship and financial support from the SKA South Africa. MJJ acknowledges support from the SKA South Africa. Opinions expressed and conclusions arrived at are those of the authors and not necessarily attributed to the SKA SA. We thank Philip Best and Roberto Maiolino for useful discussions. Thanks to Matt Prescott for his help with the optical spectra.

%------------------------------------------------------------------------------%
%
% using custom format in ADS:
%
% %z132 \\bibitem[%\2m%(y)%\3m]%{R}\n   %\8.1g,%\Y,%\q,%\V,%\p

\setlength{\labelwidth}{0pt}

\bsp

%\appendix
%\section{Table of source properties}\label{appendix}

\label{lastpage}
%------------------------------------------------------------------------------%
\end{document}